\documentclass[twocolumn,twocolappendix]{aastex631}
\usepackage{newtxtext,newtxmath}
\usepackage{multirow}
\usepackage{url}
\usepackage{amsmath}	
\usepackage{amssymb}	
\usepackage{graphicx}
\usepackage{color}
\usepackage{natbib}
\usepackage{hyperref}
\usepackage{enumitem}
\usepackage{ulem}
\setitemize{noitemsep,topsep=0pt,parsep=0pt,partopsep=0pt,leftmargin=*}
\usepackage{bm}
\usepackage{ae,aecompl}
\usepackage{color}

\renewcommand{\vec}[1]{ \bm{#1}}
\newcommand{\nvec}[1]{ \hat{\bm{#1}}}

\definecolor{turq}{rgb}{.1,.3,.5}
\definecolor{cela}{rgb}{.0,.6,.5}

\begin{document}


\title{Spectral and Polarization Properties of Coherent Radio Emission from Relativistic Magnetized Pair Plasma Shocks}

\author[0000-0003-4721-4869]{Yuanhong Qu}\thanks{E-mail: yuanhong.y.qu@helsinki.fi}
\affiliation{Department of Physics, University of Helsinki, P.O. Box 64, FI-00014 University of Helsinki, Finland}
\affiliation{Nevada Center for Astrophysics, University of Nevada, Las Vegas, NV 89154}
\affiliation{Department of Physics and Astronomy, University of Nevada Las Vegas, Las Vegas, NV 89154, USA}

\author[0000-0002-3226-4575]{Joonas N{\"a}ttil{\"a}}
\thanks{E-mail: joonas.nattila@helsinki.fi}
\affiliation{Department of Physics, University of Helsinki, P.O. Box 64, FI-00014 University of Helsinki, Finland}
\affiliation{Department of Astronomy and Columbia Astrophysics Laboratory, Columbia University, New York, NY, 10027, USA}

\author[0000-0002-1227-2754]{Lorenzo Sironi}
\thanks{E-mail: lsironi@astro.columbia.edu}
\affiliation{Department of Astronomy and Columbia Astrophysics Laboratory, Columbia University, New York, NY, 10027, USA}
\affiliation{Center for Computational Astrophysics, Flatiron Institute, 162 5th Avenue, New York, NY 10010, USA}

\begin{abstract}
Fast radio bursts (FRBs) are millisecond-duration radio transients whose emission mechanism remains an open question.
The so-called synchrotron maser instability in relativistic magnetized shocks is a leading candidate for FRBs produced outside the compact object magnetosphere, yet its spectral and polarization predictions have not been systematically explored.
We present three-dimensional particle-in-cell simulations of relativistic magnetized pair plasma shocks with upstream magnetization $\sigma > 1$ and bulk Lorentz factor $\gamma_0 = 10$, assuming a plane-parallel shock geometry and neglecting radiative cooling, and compute the spectra and polarization of the precursor maser emission for an arbitrary line of sight.
For on-axis observers aligned with the shock-propagation direction, the X-mode dominates the emission. 
The O-mode power increases at off-axis viewing angles and is primarily associated with currents parallel to the upstream background magnetic field.
Assuming a spherical shock, we find full-width-at-half-maximum fractional bandwidths of $\Delta\nu/\nu_0\simeq0.57$ and $0.82$ for $\sigma=3$ and $6$, respectively. 
For highly magnetized shocks, the intrinsic emission bandwidth accounts for most of the spectral width, while high-latitude contributions provide additional broadening.
The integrated emission remains highly linearly polarized near the spectral peak, with a nearly constant polarization angle due to the dominance of the X-mode.
Our results suggest that synchrotron maser emission can produce spectra with bandwidths and high linear polarization degrees compatible with observations of some non-repeating FRBs, and may provide testable signatures for distinguishing shock-powered FRBs from other emission mechanisms.
\end{abstract}


\section{Introduction}

Fast Radio Bursts (FRBs; for reviews, see \citet{Cordes&Chatterjee2019,petroff2019,Platts2019,Zhang2020,Lyubarsky2021,ZhangRMP}) are highly coherent millisecond-duration radio bursts detected at frequencies from $\sim100$~MHz to several GHz \citep{Lorimer2007,Thornton2013}.
Magnetars serve as the leading central-engine candidate for FRBs, supported by the detection of FRB 20200428
from the Galactic magnetar SGR 1935+2154 \citep{CHIME/FRB2020,Bochenek2020,CKLi21}.
In the magnetar framework, radiation mechanisms divide into two categories: those operating inside the magnetosphere and those operating outside the magnetosphere \citep{Zhang2020,ZhangRMP}.
The radiation site and the emission mechanism, however, remain unknown and actively debated.

Polarization and spectral observations of FRBs increasingly constrain emission mechanisms.
Most repeating FRBs are highly linearly polarized, while a small fraction exhibit significant circular polarization.
Polarization measurements encode information about both intrinsic emission mechanisms and propagation effects \citep{Qu&Zhang2023}.
We summarize the key observational features:
\begin{itemize}
\item Most repeating FRBs show high linear polarization $\Pi_L\sim 100\%$ \citep{petroff2019}.
Several repeating FRBs lack detected circular polarization, including FRB 20180301A \citep{Price2019,Luo2020nature}, FRB 20180916B \citep{Nimmo2021}, FRB 20190417A \citep{Feng2022}, FRB 20190604A \citep{Feng2022}, FRB 20190711A \citep{Day2020,PKumar2021}, and a few other CHIME repeaters \citep{Fonseca2020}.
\item A minority of bursts from some repeating FRBs exhibit significant circular polarization, with circular polarization degrees ranging from a few to tens of percent, e.g., reaching up to $\Pi_V\sim90\%$ in FRB 20201124A \citep{Jiang22,Jiang2024}, $\Pi_V\sim70\%$ in FRB 20220912A \citep{ZhangYK2023}, $\Pi_V\sim75\%$ in FRB 20240114A \citep{WangTC2026}.
For FRB 20201124A, the frequency-independent circular polarization degrees observed in some bursts cannot be explained by Faraday conversion or synchrotron / cyclotron absorption \citep{Jiang2024}.
The small and stable rotation measure (RM) of FRB 20220912A suggests a relatively clean local environment, disfavoring propagation effects as the origin of its high circular polarization \citep{ZhangYK2023,Feng2024}. 
No evidence for Faraday conversion has been found in FRB 20240114A, suggesting that its observed circular polarization is likely intrinsic to the emission mechanism \citep{Uttarkar2026,WangTC2026}. 
These observations suggest that high circular polarization may be linked to the intrinsic FRB radiation mechanism.
\item Both constant and varying polarization angles (PAs) appear in the same repeating FRB source.
Most FRB bursts exhibit nearly constant PAs and a small fraction display significant variations.
Varying PA can be classified into two forms: regular S-shaped swings and irregular swings.
A pulsar-like S-shaped PA swing has been observed in the non-repeating FRB 20221022A \citep{Mckinven2025}.
FRBs can also exhibit $\sim90^\circ$ PA jumps within individual millisecond-duration bursts, analogous to the abrupt PA jumps commonly observed in radio pulsars \citep{Manchester1975,Backer1976}. 
Such PA jumps have been detected in FRB 20201124A and attributed to orthogonal-mode switching \citep{NiuJR2024}.
\end{itemize}

Spectral features further constrain FRB emission models.
Statistical analysis of the first CHIME/FRB catalog---62 bursts from 18 repeating FRBs and 474 from non-repeating FRBs---shows that repeaters display significantly narrower frequency bandwidths than non-repeaters \citep{Pleunis2021}.
The fractional bandwidths $\Delta\nu/\nu_0$ of several active repeaters, where $\Delta\nu$ is the full width at half-maximum (FWHM) and $\nu_0$ is the central frequency, confirm this narrow-band character:
FRB 20201124A has an extremely narrow bandwidth of $\sim277$~MHz (full frequency width at 10\% of the emission peak) detected by the Five-hundred-meter Aperture Spherical radio Telescope (FAST) over 1.0--1.5~GHz \citep{ZhouDJ2022};
FRB 20220912A has $\Delta\nu/\nu_0\sim0.1$--$0.2$ \citep{ZhangYK2023};
and FRB 20230607A has $\Delta\nu/\nu_0\sim0.1$--$0.2$ \citep{ZhouDJ2025}.
In contrast to repeating FRBs, non-repeating FRBs tend to exhibit broader bandwidths \citep{Pleunis2021,CHIMECat2}.
These spectral differences suggest that repeating and non-repeating FRBs may involve distinct emission mechanisms.
Alternatively, it has been speculated that they may arise from different time-frequency evolution within a common emission framework \citep{Metzger2022}.

The synchrotron maser instability\footnote{Although \cite{Sironi2021} showed that the emission mechanism does not involve a population inversion, as originally postulated for the synchrotron maser instability, we retain this terminology, as it is widely used in the literature.} at strongly magnetized shocks in magnetar winds provides a candidate emission mechanism for FRBs \citep{Iwamoto2018,Metzger2019,Iwamoto2019,Plotnikov&Sironi2019,Beloborodov2020,Babul&Sironi2020,Iwamoto2024}.
Recent studies have investigated the formation of relativistic magnetized shocks through fast magnetosonic wave steepening in magnetar magnetospheres and the kinetic mechanisms of shock dissipation \citep{Vanthieghem&Levinson2025PRE,Vanthieghem&Levinson2025PRL}.
The synchrotron maser model predicts high linear polarization $\Pi_L\sim100\%$ dominated by the X-mode when the magnetization exceeds unity.
The O-mode is also produced, but its relative strength decreases with increasing magnetization in 3D particle-in-cell (PIC) simulations \citep{Sironi2021}.

In this paper, we investigate the spectral and polarization properties of the synchrotron maser emission model outside the magnetosphere, by performing and analyzing 3D PIC simulations of relativistic magnetized shocks.
Previous simulation-based studies focused on the line of sight (LOS) aligned with the shock propagation direction.
We go beyond this restriction by systematically measuring the emission at arbitrary viewing angles in our simulations. We then integrate the emission over the spherical shock surface to include high-latitude emission and confront the synchrotron maser predictions with FRB observations.
The paper is organized as follows.
Section~\ref{sec:simulation set up} presents the methods and numerical set-up.
Section~\ref{sec:dispersion} discusses the dispersion relations of X- and O-modes.
Section~\ref{sec:individual} describes simulation results in the downstream (simulation) rest frame, and Section~\ref{sec:integrated} covers the viewing-angle-integrated case.
In Section~\ref{sec:analytical}, we analytically investigate the spectral broadening caused by the high-latitude effect and its dependence on the magnetization.
Section~\ref{sec:conclusion} summarizes our conclusions and discussions.
The convention $Q=10^nQ_{n}$ in cgs units is adopted.

\begin{figure*}
\centering
\includegraphics[width=16cm]{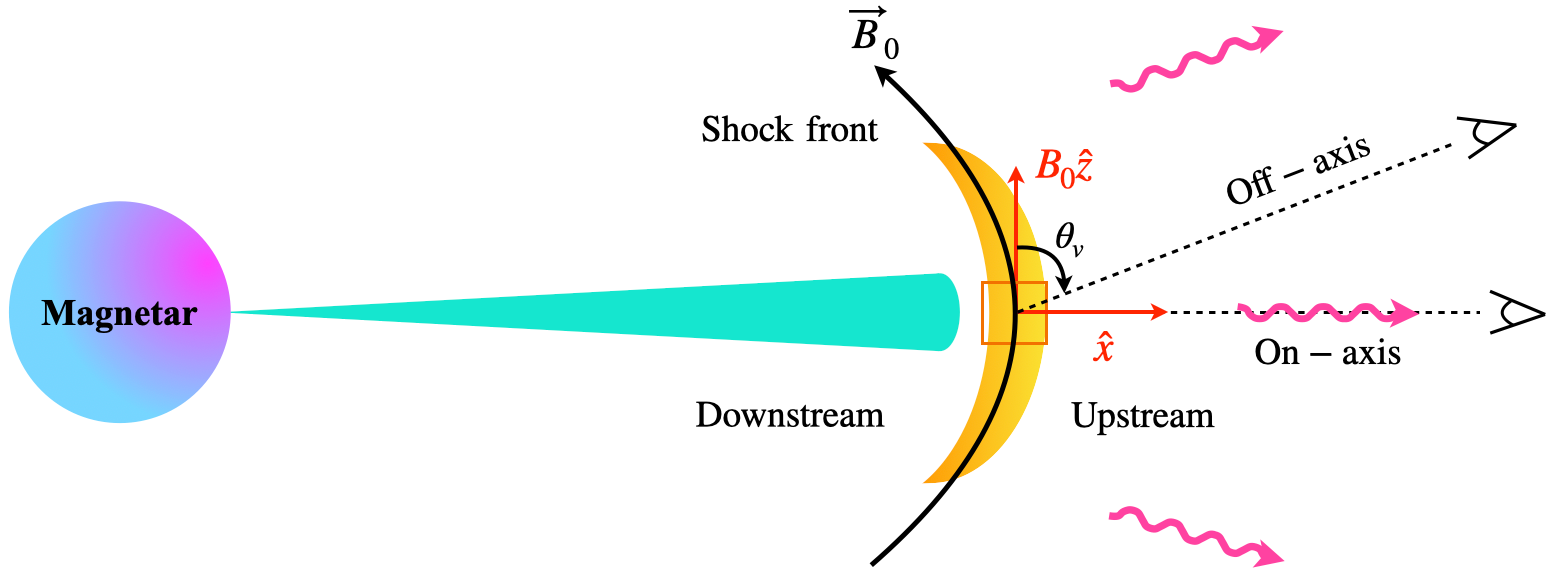}
    \caption{Geometry of the relativistic magnetized pair plasma shock simulation. The shock propagates radially outward, and the upstream magnetic field $\vec B_0$ is assumed to be toroidal and thus perpendicular to the shock propagation direction. 
    Coherent radio waves (purple wigglers) originate from the downstream region behind the shock front. The viewing angle $\theta_v$ measures the angle between the observer's LOS and the local background magnetic field; $\theta_v=90^\circ$ corresponds to the on-axis view along the shock propagation direction, while smaller $\theta_v$ sample off-axis emission.
    The two LOSs equivalently correspond to different portions of the shock surface seen by a single observer, with small $\theta_v$ corresponding to high-latitude emission.}
    \label{fig:shock_geo}
\end{figure*}

\begin{figure}
\hspace*{-2mm}
\includegraphics[width=86mm]{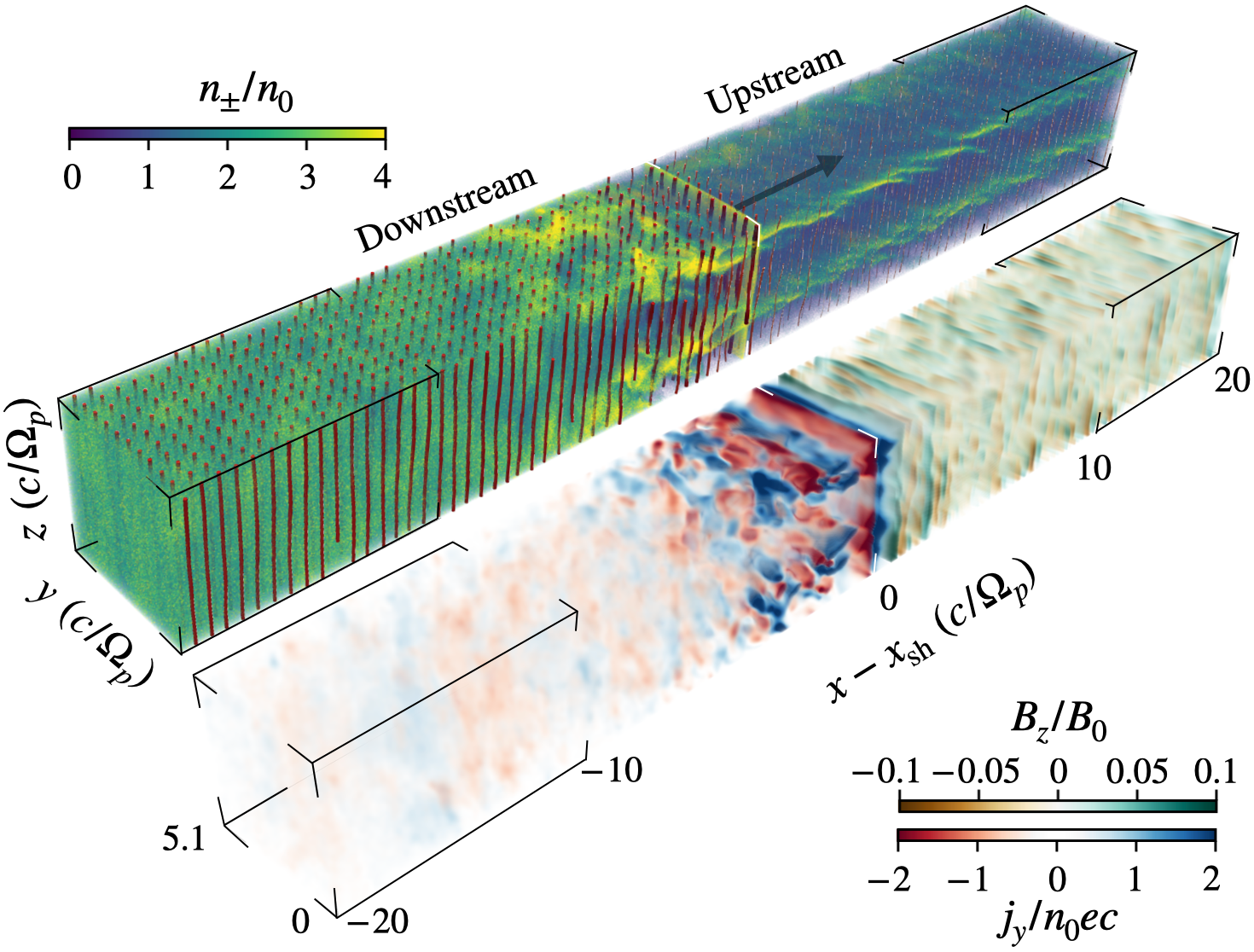}
    \caption{3D rendering of the shock structure, showing density (top), magnetic field $B_z/B_0$ (bottom, in the upstream) and current density $j_y/n_0ec$ (bottom, in the downstream) from a simulation with $\gamma_0=10$ and $\sigma=3$ at $\Omega_p t=2000$. 
    The $x$ coordinate is measured with respect to the shock location $x_{\rm sh}$ in units of $c/\Omega_p$. The upstream is at $x-x_{\rm sh}>0$ and the downstream at $x-x_{\rm sh}<0$. 
    The arrow in the top panel indicates the direction of shock propagation.
    Magnetic field lines are shown with red tubes with thickness proportional to $|\vec{B}|$.
    The oscillating electric currents (red and blue) excite the electromagnetic radiation (green and brown) to the shock upstream. Upstream plasma filamentation, which causes the O-mode component, is visible as stripes in the plasma density.
    }
    \label{fig:shock_structure}
\end{figure}

\section{Simulation methods and set-up}\label{sec:simulation set up}

We use the electromagnetic PIC code \textsc{runko} to perform 3D shock simulations in the downstream frame \citep{Nattila2022}, which we refer to as the simulation frame.
The basic configuration of our simulations is illustrated in Figure~\ref{fig:shock_geo}.
The upstream flow is a cold pair plasma drifting in the $-\nvec{x}$ direction with a bulk Lorentz factor $\gamma_0=10$.
The pair plasma in the upstream carries a frozen-in magnetic field $B_0$ oriented along the $z$-axis.
The upstream magnetization is defined as the ratio of Poynting to kinetic energy flux
\begin{equation}
\sigma=\frac{B_0^2}{4\pi\gamma_0 n_0 m_e c^2}=\frac{B_0'^2}{4\pi n_0' m_e c^2}=\frac{\Omega_B^2}{\Omega_p^2}=\frac{\omega_B'^2}{\omega_p'^2},
\end{equation}
where primed quantities are measured in the rest frame of the upstream. 
The gyration frequency and the plasma frequency in the simulation frame are defined as
\begin{equation}
\Omega_B=\frac{eB_0}{\gamma_0 m_e c}, \ \Omega_p=\left(\frac{4\pi e^2n_0}{\gamma_0m_e}\right)^{1/2},
\end{equation}
and in the upstream rest frame as
\begin{equation}
\omega_B'=\frac{eB_0'}{m_ec}, \ \omega_p'=\left(\frac{4\pi e^2n_0'}{m_e}\right)^{1/2},
\end{equation}
where $B_0'=B_0/\gamma_0$ for a perpendicular shock and $n_0'=n_0/\gamma_0$ is the
total number density of upstream positrons and electrons, $m_e$ is the electron mass, and $e$ is the elementary electric charge.
We also note $\Omega_B=\omega_B'$ and $\Omega_p=\omega_p'$.

In the simulation, the upstream moves with velocity $-\beta_0 c\nvec{x}$ and the background magnetic field $\vec B_{0}$ is along the $z$-axis. 
Thus a motional electric field $E_{y,0}=-\beta_0B_{z,0}$ is required in the rest frame of the wall (simulation frame) to ensure that the electric field in the rest frame of the upstream vanishes. 
The particle flow is reflected at the left wall and the shock front is formed to steadily propagate along the $+\hat{\boldsymbol x}$ direction.
Figure~\ref{fig:shock_structure} shows the shock structure from a simulation when the system has reached a quasi-steady state, corresponding to the local region indicated by the red box in Figure~\ref{fig:shock_geo}.

We resolve the upstream plasma skin depth with $c/\Omega_p=25\Delta$, where $\Delta$ is the spatial grid size, on a grid of $N_x\times N_y\times N_z=65536\times128\times128$ cells.
The transverse box size is $L_y=L_z=5.12c/\Omega_p$, sufficient to resolve the filamentation modes \citep{Sironi2021,Sobacchi2025}.
This resolution ensures that the particle gyroradius $r_g\simeq \sigma^{-1/2}c/\Omega_p$ is well resolved up to the largest magnetization $\sigma=6$ studied here.
The simulation time step is $c\Delta t=0.5\Delta$, corresponding to $\Delta t=0.02\Omega_p^{-1}$.
We initialize $N_{\rm ppc}=2$ particles per cell per species in the upstream plasma with a thermal spread $k_BT_0/(m_ec^2)=10^{-4}$.
Numerical Cherenkov radiation is mitigated by using a 3D-generalization of the \citet{blinne2018} field solver (N\"attil\"a, in prep.).
Particle momentum is updated with a \citet{higuera2017} pusher and current deposited with a ZigZag scheme \citep{umeda2003}.
We filter the current with 8 Binomial filter passes.
Each simulation runs for several thousand $\Omega_p^{-1}$, long enough for both the X- and O-mode precursor waves to reach a quasi-steady state with nearly constant amplitude.

The adopted $N_{\rm ppc}=2$ per species in this paper is motivated by previous numerical studies of relativistic magnetized pair shocks. In 2D simulations, \cite{Plotnikov&Sironi2019} tested particle loadings between 2 and 32 particles per cell per species and found no significant differences (see their Appendix~B). 
In 3D simulations, \cite{Sironi2021} reported converged results with as few as one particle per cell (see their Supplementary Material).

\section{Dispersion relations of X-mode and O-mode in the highly magnetized plasmas}\label{sec:dispersion}

In the highly magnetized regime considered here, both X- and O-mode waves approach the dispersion relation of electromagnetic waves in vacuum.
This allows us to directly relate the wavenumber spectra measured in the PIC simulations to the corresponding frequency spectra.

In the simulation frame, the peak frequency of the precursor emission for $\sigma\gtrsim1$ is \citep{Plotnikov&Sironi2019}
\begin{equation}
\omega\simeq \xi\sqrt{\sigma}\Omega_p,
\end{equation}
with $\xi\simeq3$ found in 1D PIC simulations.
The refractive indices of O- and X-modes can be written as
\begin{equation}
n_{\rm O}^2
=1-\frac{\Omega_p^2}{\omega^2}
\simeq1-\frac{1}{\xi^2\sigma},
\end{equation}
and
\begin{equation}
n_{\rm X}^2
=1-\frac{\Omega_p^2}{\omega^2-\Omega_B^2}
\simeq1-\frac{1}{(\xi^2-1)\sigma}.
\end{equation}
For $\sigma\gg1$, both modes satisfy
\begin{equation}
n_{\rm O}\simeq n_{\rm X}\simeq1.
\end{equation}
We conclude that both X- and O-modes have nearly identical dispersion relations and propagate approximately as vacuum electromagnetic waves.
Thus their $\omega$-spectra can be directly identified with their $k$-spectra in the simulation frame.

\begin{figure*}
\setlength{\tabcolsep}{0pt}
\begin{center}
\begin{tabular}{ll}
\resizebox{94mm}{!}{\includegraphics[]{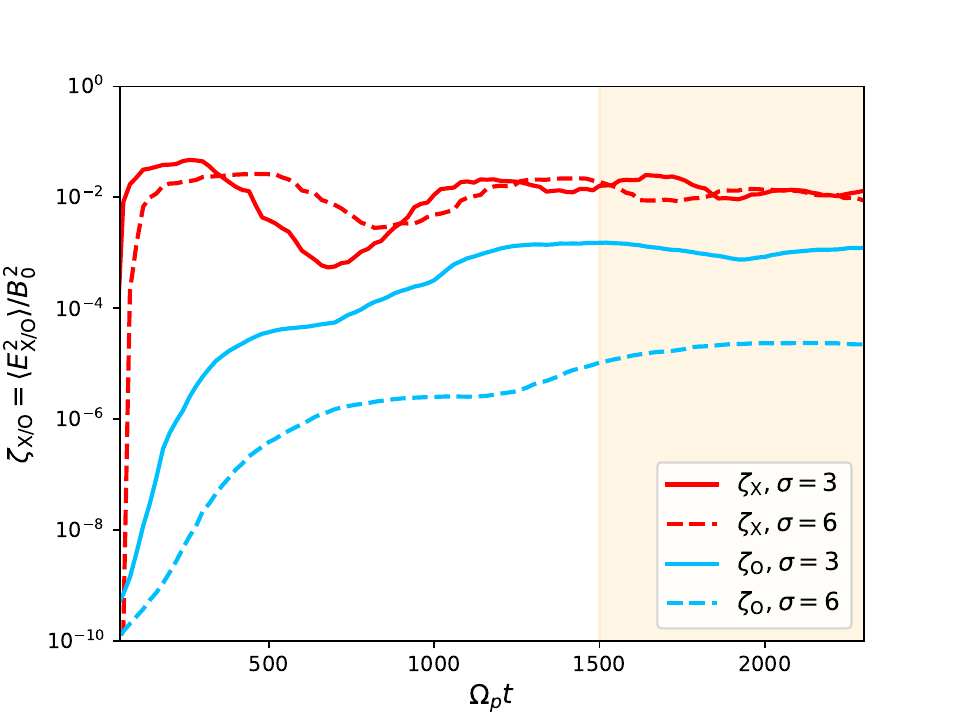}}&
\resizebox{94mm}{!}{\includegraphics[]{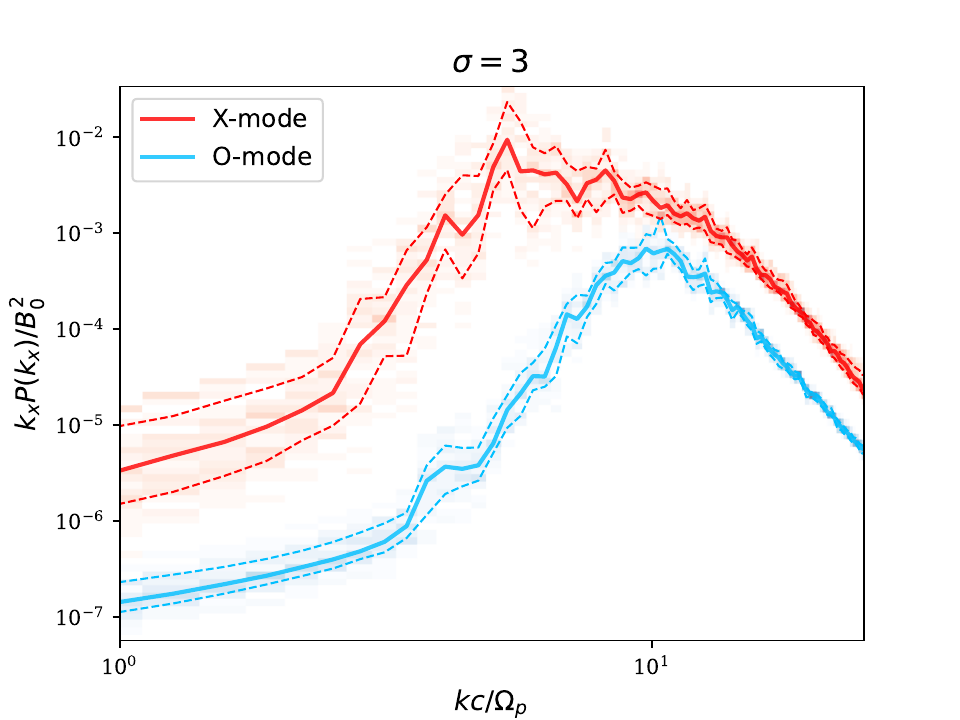}}
\end{tabular}
\caption{Left panel: Time evolution of the Poynting flux carried by X- and O-modes, $\zeta_{\rm X/O}=\langle E_{\rm X/O}^2\rangle/B_0^2$, for upstream magnetizations $\sigma=3$ (solid lines) and $\sigma=6$ (dashed lines). The precursor wave energy was extracted from a $30c/\Omega_p$ wide slab located at $5c/\Omega_p < x-x_{\rm sh} < 35c/\Omega_p$.
The shaded region marks the quasi-steady time interval $1500\leq\Omega_p t\leq2300$ used for the right panel and for all our subsequent analysis.
Right panel: temporal evolution of the 1D X- and O-mode power spectra for $\sigma=3$ in the downstream frame.
The solid curves show the median spectra over the quasi-steady time interval $1500 \leq \Omega_pt \leq 2300$, and the dashed curves show the 16th and 84th percentiles over the same time interval.
The shaded regions indicate the temporal variation of the spectra.
Here $P(k_x)=\int P_{\rm 3D}(k_x,k_y,k_z)dk_ydk_z$ and $P_{\rm 3D}$ is the 3D power spectral density.
}
\label{fig:kspec_all_sig3}
\end{center}
\end{figure*}

\section{Simulation results}\label{sec:individual}

In this section, we present the spectral and polarization properties of coherent emission from relativistic magnetized pair plasma shocks across a range of viewing angles.
All quantities are measured in the downstream frame.
The analysis uses the Fourier-transformed electric fields described in Appendix \ref{app:polarization}.
The signal is attenuated by a Hamming window function in the $\nvec{x}$-direction to account for the non-periodic $x$ boundaries.
The spectra are computed from the upstream region $5c/\Omega_p < x-x_{\rm sh} < 35c/\Omega_p$ ahead of the shock front.
The corresponding minimum wavenumber is $k_{\min}c/\Omega_p \sim 0.2$, and we restrict our analysis to $kc/\Omega_p \geq 1$.

The simulation domain represents a finite patch of the shock front (red box in Figure~\ref{fig:shock_geo}), whose transverse extent (perpendicular to the $x$-axis) is much smaller than the curvature radius of a realistic shock in the magnetar wind region.
Thus each PIC simulation models one local patch of the shock front.
Different viewing angles of a single simulated shock patch correspond to different patches of a spherical shock viewed by a single observer.
We defer the integration of emission from different latitudes of the shock surface to Section~\ref{sec:integrated}.

\subsection{Radiation spectra}

We decompose the radiation from the relativistic shock into the X-mode and O-mode eigenmodes perpendicular to the LOS and compute their spectra for arbitrary viewing angles.
The LOS unit vector is $\nvec{n}=\sin\theta_v\cos\phi_v\nvec{x}+\sin\theta_v\sin\phi_v\nvec{y}+\cos\theta_v\nvec{z}$.
The spherical-shock geometry adopted here is azimuthally symmetric, allowing us to choose the local coordinate system such that the LOS lies in the $x-z$ plane ($\phi_v=0$) without loss of generality, i.e. $\nvec{n}=\sin\theta_v\nvec{x}+\cos\theta_v\nvec{z}$, which contains both the shock normal and the background magnetic field $\vec{B}_0$.
The two orthogonal complex Fourier-space electric field amplitudes from Appendix~\ref{app:polarization} simplify to
\begin{equation}
\widetilde{E}_1=\widetilde{E}_y,
\end{equation}
for the X-mode, and
\begin{equation}
\begin{aligned}
\widetilde{E}_2&={\widetilde{E}}_x\cos\theta_v-{\widetilde{E}}_z\sin\theta_v,
\end{aligned}
\end{equation}
for the O-mode.
The spectral intensities of the X-mode and O-mode are then
\begin{equation}
|\widetilde E_{\rm X}|^2=|\widetilde E_y|^2,
\end{equation}
and
\begin{equation}
|\widetilde E_{\rm O}|^2=|{\widetilde{E}}_x\cos\theta_v-{\widetilde{E}}_z\sin\theta_v|^2.
\end{equation}

Before analyzing the viewing-angle dependence of the emission, we first examine the temporal stability of the spectra. 
The left panel of Figure~\ref{fig:kspec_all_sig3} shows the time evolution of the Poynting flux carried by X- and O-modes, both of which reach a quasi-steady state at late times $1500 \leq \Omega_p t \leq 2300$.
The right panel shows the X- and O-mode spectra at different times over this interval.
The spectral shape around the emission peak remains stable, although there are relatively large temporal variations at low wavenumbers.
We first analyze a representative snapshot for each simulation in the following.

\begin{figure*}
\begin{center}
\begin{tabular}{ll}
\resizebox{86mm}{!}{\includegraphics[]{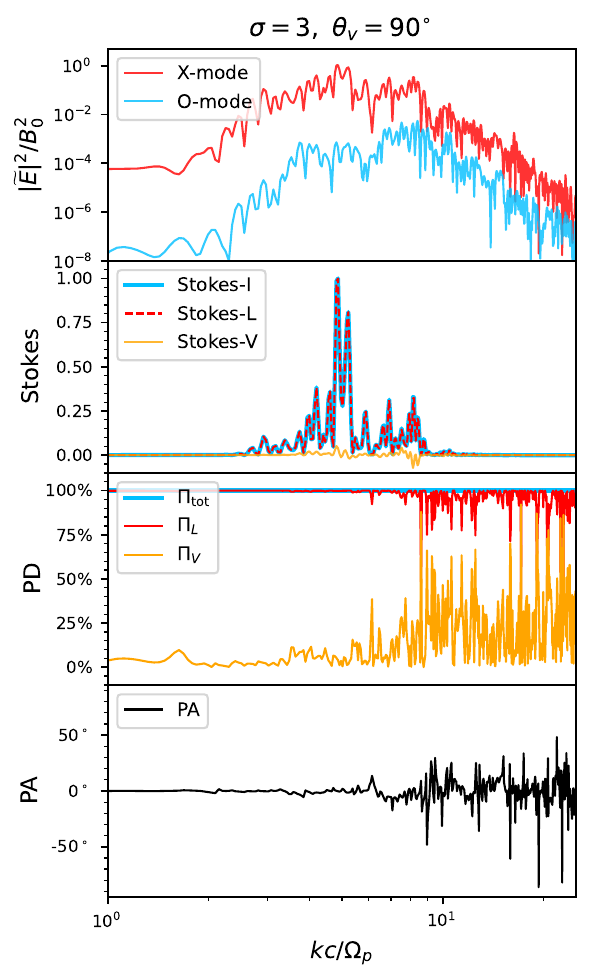}}&
\resizebox{86mm}{!}{\includegraphics[]{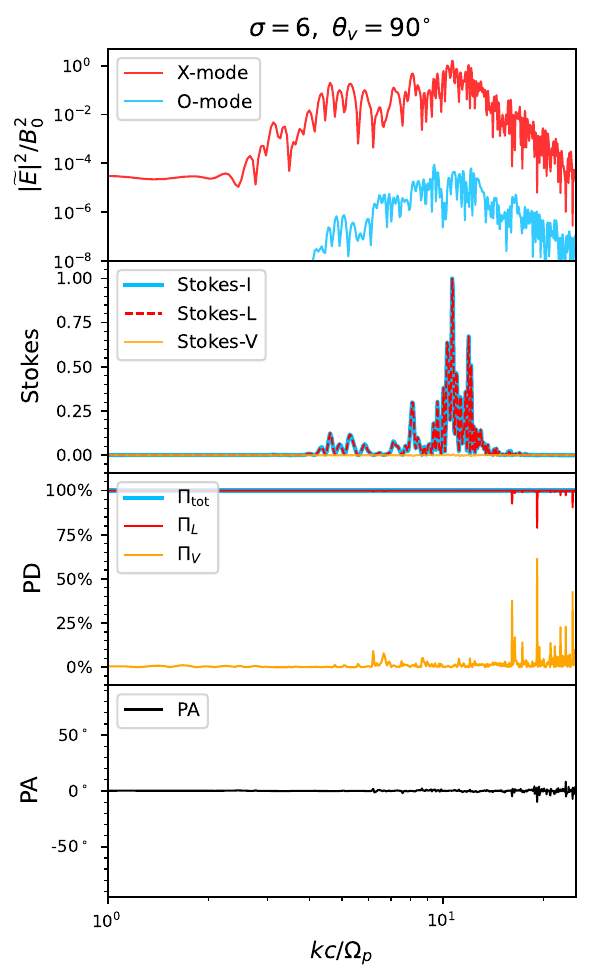}}
\end{tabular}
\caption{Radiation spectra, Stokes parameters, polarization degree (PD), and polarization angle (PA) as a function of $kc/\Omega_p$ for on-axis observer ($\theta_v=90^\circ$) for $\sigma=3$ (left panel) and $\sigma=6$ (right panel) at $\Omega_p t=1900$ in the downstream frame. 
These results are extracted directly from the PIC simulations, which adopt $\gamma_0=10$.
The X-mode dominates the entire emission band, producing a high degree of linear polarization near the spectral peak; circular polarization grows at high frequencies where the O-mode becomes comparable to the X-mode.}
\label{fig:single_on_axis}
\end{center}
\end{figure*}

For the on-axis observer with $\theta_v=90^\circ$, the X-mode corresponds to $\widetilde{E}_y$ and the O-mode to $\widetilde{E}_z$.
The first row of Figure~\ref{fig:single_on_axis} shows the X- and O-mode spectra for $\sigma=3$ and $\sigma=6$.
The X-mode dominates across the entire emission band.
In highly magnetized shocks, the pair plasma motion is confined to the $x$--$y$ plane.
The current $j_z$ arises from filamentation of the upstream plasma, which bends the magnetic field when entering the shock front.
As $\sigma$ increases, the downstream particle motion becomes increasingly confined to the $x-y$ plane, reducing the current $j_z$ that generates the O-mode emission and leading to a high degree of linear polarization. The O-mode and X-mode spectra overlap at high wavenumbers. These results agree with \cite{Sironi2021}.

\begin{figure*}
\begin{center}
\setlength{\tabcolsep}{-2pt}
\begin{tabular}{lll}
\resizebox{61mm}{!}{\includegraphics[]{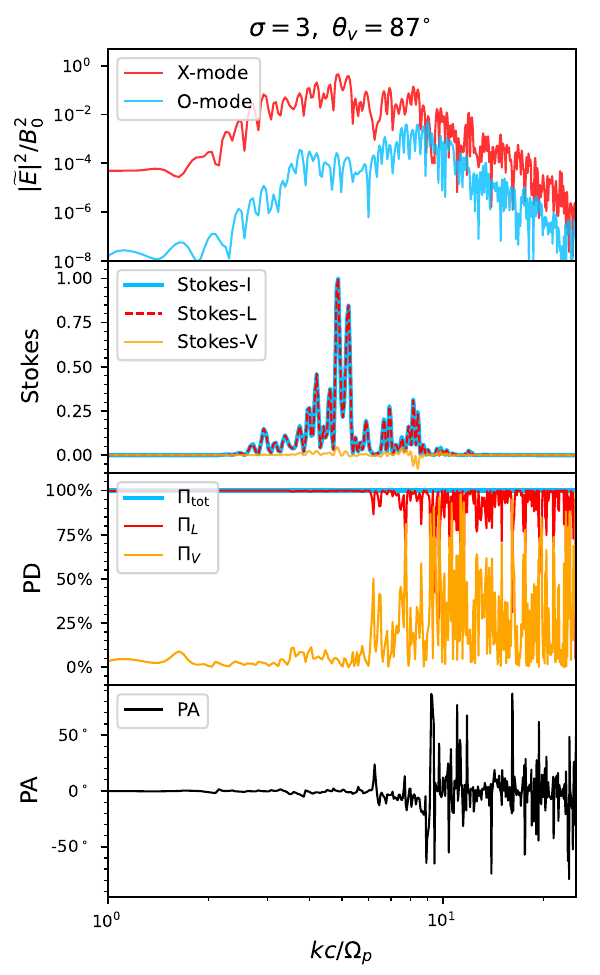}}&
\resizebox{61mm}{!}{\includegraphics[]{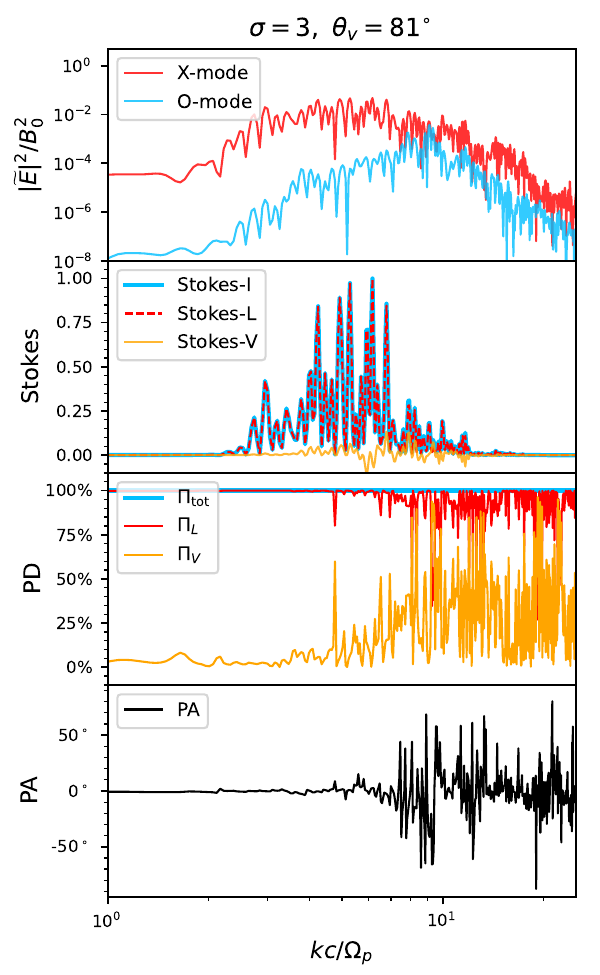}}&
\resizebox{61mm}{!}{\includegraphics[]{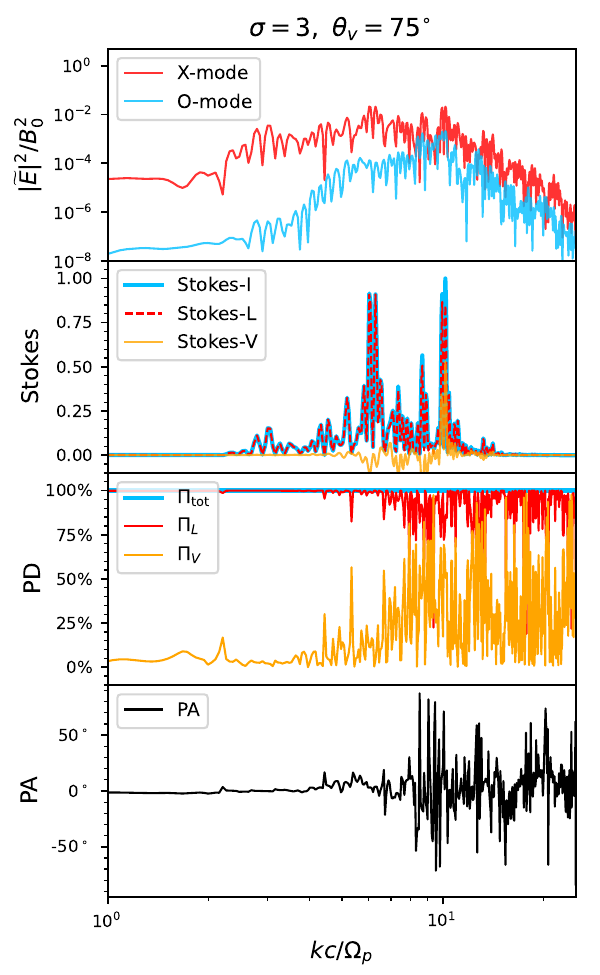}}\\
\resizebox{61mm}{!}{\includegraphics[]{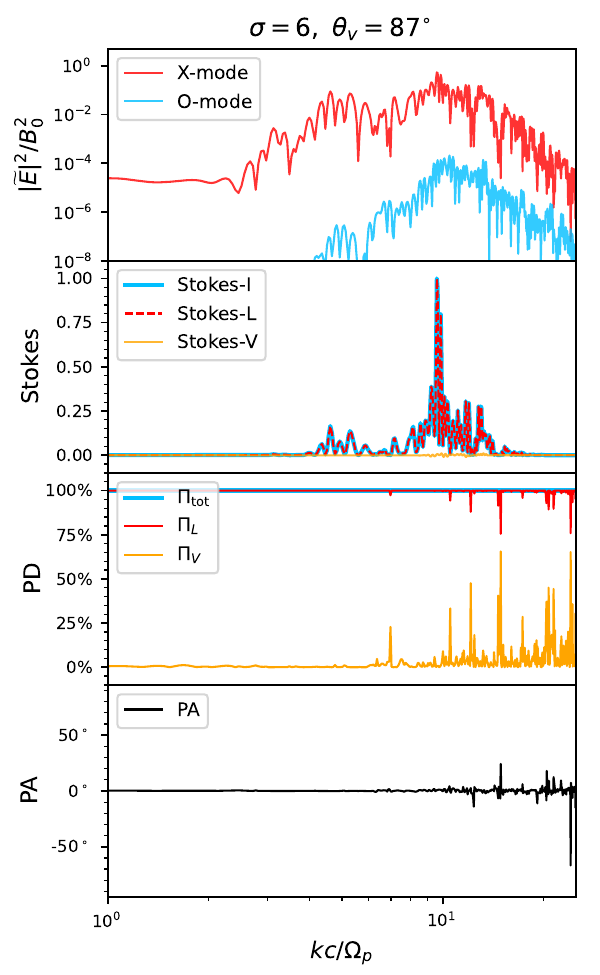}}&
\resizebox{61mm}{!}{\includegraphics[]{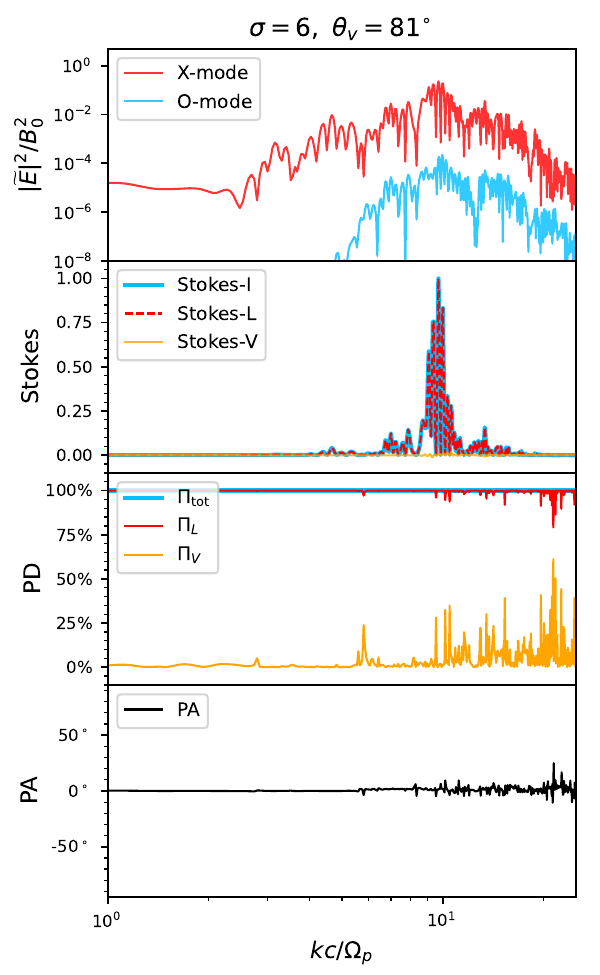}}&
\resizebox{61mm}{!}{\includegraphics[]{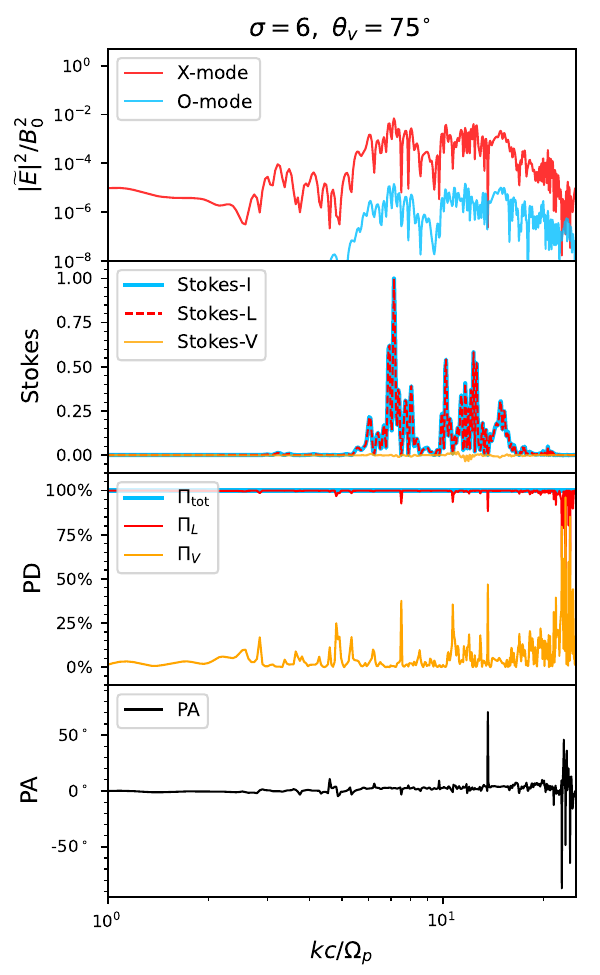}}
\end{tabular}
\caption{Radiation spectra, Stokes parameters, PD, and PA as a function of $kc/\Omega_p$ for off-axis viewing angles for $\sigma=3$ and $\sigma=6$ at $\Omega_p t=1900$ in the downstream frame. As $\theta_v$ decreases from $90^\circ$, the O-mode contribution grows at high frequencies, reducing the linear polarization degree and introducing circular polarization. The total emission power decreases with increasing misalignment.
}
\label{fig:individual_arbitrary} 
\end{center}
\end{figure*}

For different LOSs, we normalize both X-mode and O-mode spectra by the frequency-integrated on-axis X-mode emission.
Figure~\ref{fig:individual_arbitrary} shows the X- and O-mode spectra for a range of off-axis viewing angles.
As the LOS becomes increasingly misaligned with respect to the shock normal, the O-mode emission approaches the X-mode, particularly at high frequencies.
As $\theta_v$ decreases, an increasing fraction of the $x$-directed current is projected onto the plane perpendicular to the LOS, providing an additional contribution to the O-mode emission. However, this contribution remains subdominant, as the O-mode emission is primarily associated with the electric current parallel to the background magnetic field (i.e., along the $z$-axis).

To isolate the contributions from different radiation field components, we first set $\widetilde{E}_z = 0$, in which case the X-mode then dominates at all viewing angles.
We next set $\widetilde{E}_x = 0$ and find that the relative X- and O-mode intensities remain nearly identical to those in Figure~\ref{fig:single_on_axis} and \ref{fig:individual_arbitrary}, indicating that the dominant O-mode emission is associated with the current along the $z$-axis. 
In the limiting case $\theta_v\rightarrow0$, both modes are strongly suppressed as the LOS lies outside the radiation cone.

\subsection{Polarization properties}\label{subsec:polarization}

The four Stokes parameters characterize the polarization of a quasi-monochromatic electromagnetic wave \citep{Rybicki&Lightman1979}:
\begin{equation}\label{eq:Stokes}
\begin{aligned}
&I=\frac{1}{2}(\widetilde{E}_1^{*}\widetilde{E}_1+\widetilde{E}_2^{*}\widetilde{E}_2), \ &&Q=\frac{1}{2}(\widetilde{E}_1^{*}\widetilde{E}_1-\widetilde{E}_2^{*}\widetilde{E}_2),\\
&U={\rm Re}(\widetilde{E}_1^{*}\widetilde{E}_2), \ &&V={\rm Im}(\widetilde{E}_1^{*}\widetilde{E}_2),
\end{aligned}
\end{equation}
where $\widetilde{E}_1$ and $\widetilde{E}_2$ are the complex Fourier-space electric field amplitudes of the two linearly polarized eigenmodes perpendicular to the LOS, the superscript $``*"$ denotes the complex conjugation, $I$ is the total intensity, $Q$ and $U$ characterize the magnitude and orientation of linear polarization, and $V$ measures circular polarization. The degrees of linear, circular, and total polarization are $\Pi_{L}=(Q^2+U^2)^{1/2}/I$, $\Pi_{V}=|V|/I$, and $\Pi_{\rm tot}=(Q^2+U^2+V^2)^{1/2}/I$, respectively, all satisfying $\leq1$ by definition\footnote{For emission from a single-patch at a given time, the Stokes parameters are constructed directly from the complex Fourier-space electric-field amplitudes at each frequency, and hence the total polarization degree satisfies $\Pi_{\rm tot}=1$ by construction.}.
The polarization angle (PA) is defined as
\begin{equation}
{\rm PA}=\frac{1}{2}\arctan\frac{U}{Q}.
\end{equation}

For on-axis observers ($\theta_v = 90^\circ$), the emission is strongly linearly polarized with a nearly flat PA near the peak frequency.
At higher frequencies, the O-mode contribution grows comparable to the X-mode, producing significant circular polarization and a frequency-dependent PA.

For off-axis viewing angles, the increasing O-mode contribution modifies the polarization properties. 
As $\theta_v$ decreases from $90^\circ$, the linear polarization degree decreases and the circular polarization becomes increasingly prominent, particularly at frequencies where the X- and O-mode powers become comparable. 
Yet, around the spectral peak the X-mode remains dominant, and the emission retains a high degree of linear polarization with a nearly constant PA.

\subsection{Time-averaged spectral and polarization properties}

\begin{figure*}
\begin{center}
\setlength{\tabcolsep}{-2pt}
\begin{tabular}{lll}
\resizebox{61mm}{!}{\includegraphics[]{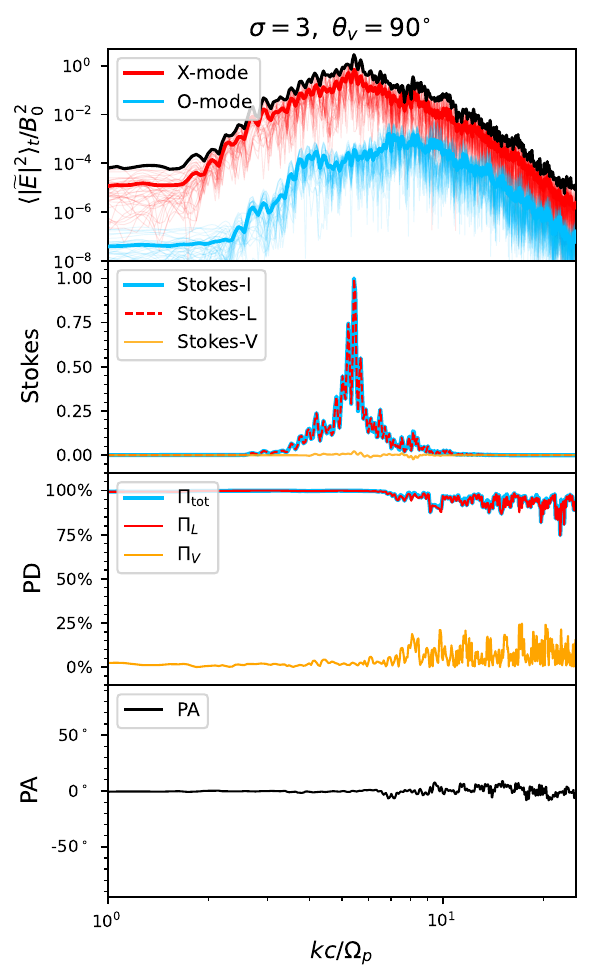}}&
\resizebox{61mm}{!}{\includegraphics[]{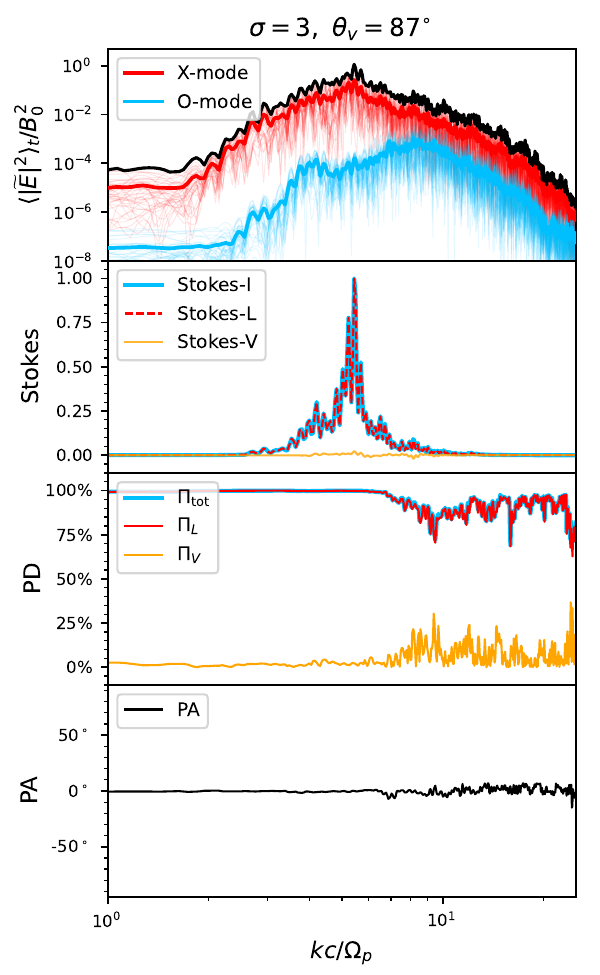}}&
\resizebox{61mm}{!}{\includegraphics[]{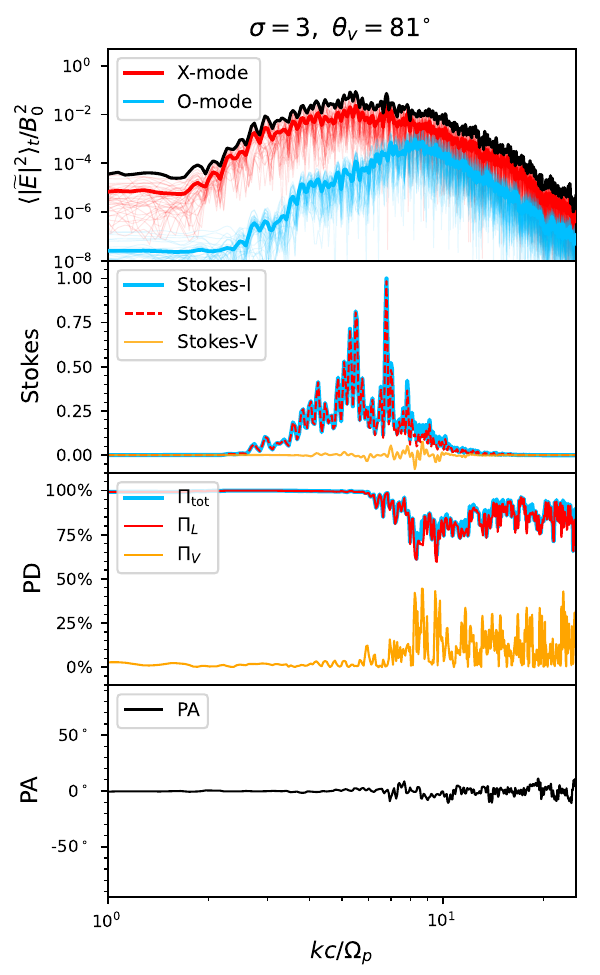}}\\
\resizebox{61mm}{!}{\includegraphics[]{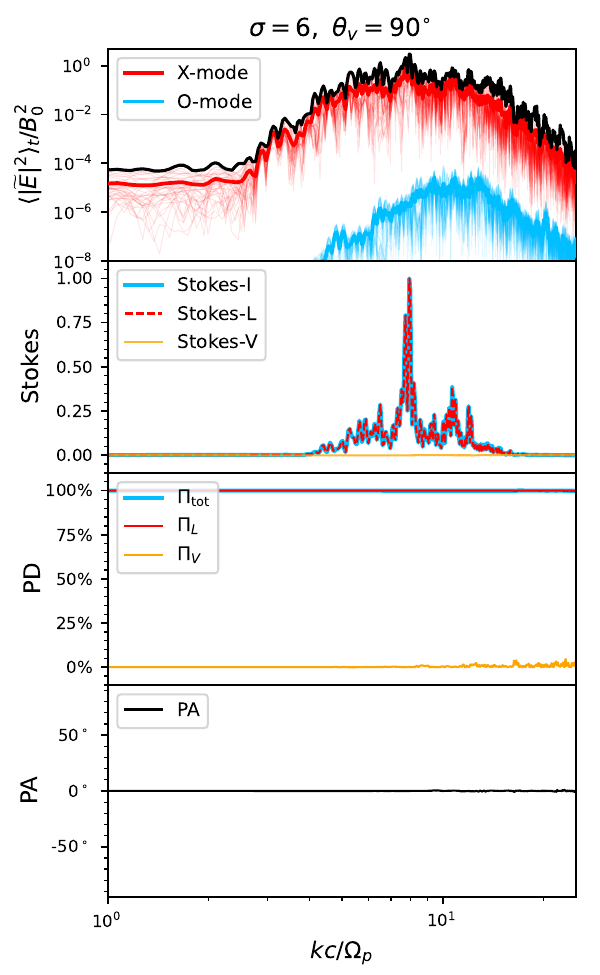}}&
\resizebox{61mm}{!}{\includegraphics[]{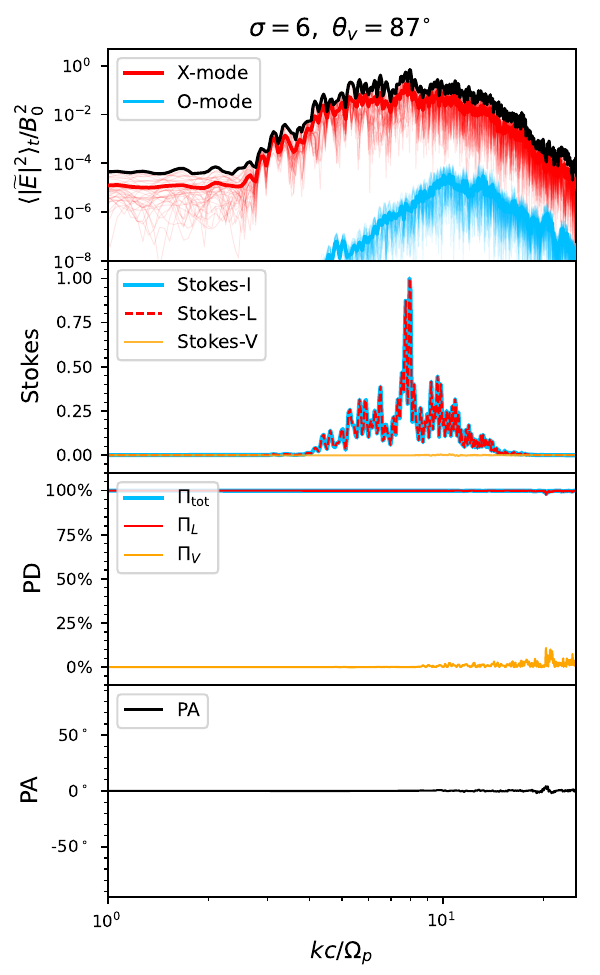}}&
\resizebox{61mm}{!}{\includegraphics[]{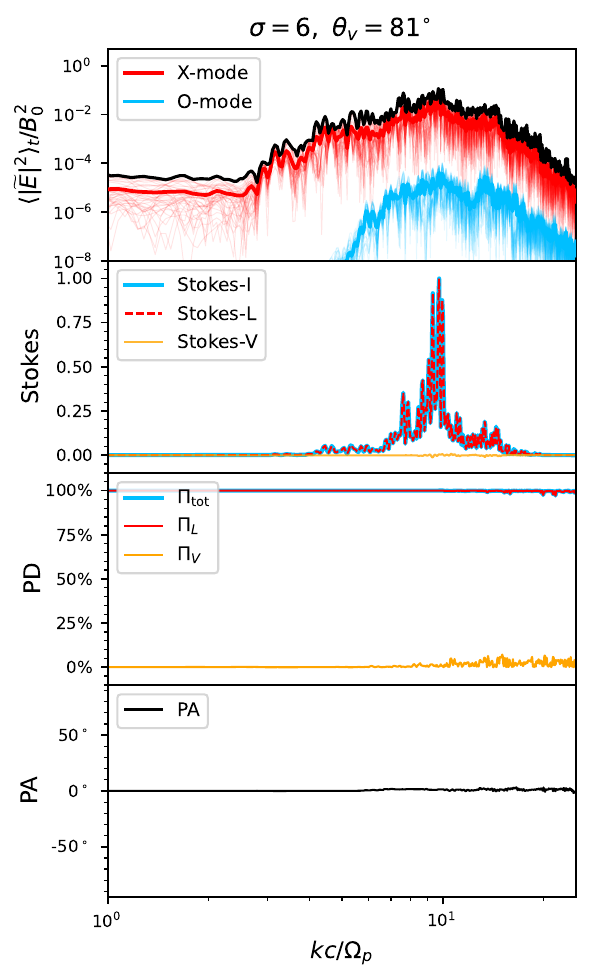}}
\end{tabular}
\caption{Time-averaged radiation spectra, Stokes parameters, PD, and PA as a function of $kc/\Omega_p$ for on-axis and off-axis viewing angles for $\sigma=3$ and $\sigma=6$ in the downstream frame. The faint curves in the upper panels show spectra from individual snapshots during the quasi-steady interval $1500\leq\Omega_p t\leq2300$ and the solid curves show their time averages.
The Stokes parameters are averaged over the same time interval, and the polarization degrees and PA are calculated from the time-averaged Stokes parameters.
The black curve shows the maximum combined power of the X- and O-modes at each wavenumber during the quasi-steady time interval.
}
\label{fig:individual_arbitrary_timeavg} 
\end{center}
\end{figure*}

To assess the temporal robustness of the spectral and polarization properties, we repeat the spectra and polarization analyses over $N_t=41$ snapshots in the quasi-steady interval $1500\leq \Omega_p t\leq2300$.
For each LOS, the X- and O-mode spectra are computed for every snapshot and then averaged as
\begin{equation}
\left\langle |\widetilde{E}_{\rm X,O}|^2 \right\rangle_t=\frac{1}{N_t}\sum_{j=1}^{N_t}|\widetilde{E}_{\rm X,O}(t_j)|^2.
\end{equation}
The Stokes parameters are averaged as
\begin{equation}
\langle S\rangle_t=
\frac{1}{N_t}\sum_{j=1}^{N_t} S(t_j), \ S\in\{I,Q,U,V\}.
\end{equation}
The time-averaged polarization degrees and PA are calculated from $\langle I\rangle_t$, $\langle Q\rangle_t$, $\langle U\rangle_t$, and $\langle V\rangle_t$ using the definitions in Section~\ref{subsec:polarization}.

Figure~\ref{fig:individual_arbitrary_timeavg} shows the time-averaged spectral and polarization properties. 
The faint curves in the spectral panels show the individual snapshots and the solid curves show the corresponding time averages.
The broad spectral envelope remains stable throughout the quasi-steady interval.
For both $\sigma=3$ and $\sigma=6$, the X-mode remains dominant around the spectral peak, and the relative contribution of the O-mode increases as $\theta_v$ decreases, particularly at high wavenumbers.
The time-averaged emission remains predominantly linearly polarized around the spectral peak since the X-mode is dominant, with a nearly constant PA.
At higher wavenumbers, the O-mode contribution becomes more important, the circular polarization degrees increase and the PAs exhibit larger variations.
Since the Stokes parameters are averaged before computing the polarization degrees, depolarization can occur, and the total polarization degree could be below unity.

\section{Integration Over a Spherical Shock}\label{sec:integrated}

Consider a relativistic spherical shell propagating with Lorentz factor $\Gamma$ in the lab frame, which corresponds to the relative Lorentz factor $\gamma_0$ between the upstream and downstream frames in the simulation.
The emission radius of the relativistic shock can be estimated as
\begin{equation}
r \simeq 2c\Gamma^2 \Delta t_{\rm FRB}
\sim (10^{12}\ {\rm cm}) \ \Gamma_2^2 \Delta_{\rm FRB,-3},
\label{eq:emission_radius}
\end{equation}
where $\Delta_{\rm FRB}$ is the typical time duration of FRBs.
This estimate assumes that the emitting region is causally connected over the angular scale $\sim 1/\Gamma$, i.e., that the shock front is sufficiently smooth and does not exhibit corrugation on angular scales smaller than $1/\Gamma$ (e.g., \cite{Demidem2023,Bresci2023}).
Equivalently, any perturbations to the shock surface have a typical angular scale larger than the emission cone of angular size $1/\Gamma$. 
The shock can therefore be approximated as spherical.
In performing the integration, each surface element is treated as a local planar shock patch with the same magnetic field geometry as in our PIC simulations, with the upstream magnetic field perpendicular to the local shock normal. 
The choice of the magnetic field direction is motivated by the magnetic geometry of a radially expanding magnetar wind. 
Outside the light cylinder of the magnetar, the radial and toroidal magnetic field components scale approximately as $B_r\propto r^{-2}$ and $B_\phi\propto r^{-1}$, respectively. 
The magnetic field is expected to be predominantly toroidal at the emission radius estimated in Equation~(\ref{eq:emission_radius}), and so approximately perpendicular to the local radial shock normal.
Through relativistic aberration, different locations on the spherical surface correspond to different viewing angles in the downstream frame sampled by the PIC simulations (see Figure~\ref{fig:shock_geo}).
An observer receives radiation not only from the patch of the shock surface moving directly along the LOS but also from higher-latitude regions whose velocities are misaligned with respect to the LOS.
We present the spectra and polarization properties obtained by integrating the emission over the spherical shock front.

Two distinct modes of superposition may arise: coherent and incoherent.
In coherent superposition, wave trains overlap along the LOS, allowing the electric fields to be added linearly before computing the Stokes parameters.
Incoherent superposition occurs when radiation arriving from different LOS originates from two spatially distinct emission regions.
The observer detects the two waves independently, so the Stokes parameters (rather than the electric fields) should be added linearly.
The phase coherence between radiation from different surface elements is controlled by the difference in their propagation path length to the observer.
At the largest observable latitude, corresponding to an angular offset of order $\sim1/(\Gamma\sqrt{\sigma})$ \citep{Babul&Sironi2020}, the characteristic path-length difference in the observer frame is $L_{\rm coh}\sim r/(\Gamma^2\sigma)$. 
If this path-length difference is smaller than or comparable to the observed radio wavelength, i.e., $L_{\rm coh}\lesssim \lambda_{\rm FRB}$, the emission from spatially separated points on the shell can add coherently.
However, for the $1/(\Gamma\sqrt{\sigma})$ radiation cone, the length scale $L_{\rm coh}$ is much larger than the FRB wavelength, i.e.,
\begin{equation}
L_{\rm coh}\simeq\frac{r}{2\Gamma^2\sigma}\sim (10^{7} \ {\rm cm}) \ \Delta t_{\rm FRB,-3}\left(\frac{\sigma}{3}\right)^{-1}\gg \lambda_{\rm FRB},
\end{equation}
where $\sigma=3$ is adopted, $\Delta t_{\rm FRB}$ and $\lambda_{\rm FRB}$ are the typical duration and wavelength of FRBs, respectively.
We define the critical opening angle $\theta_c$ by the condition that the propagation path-length difference is comparable to the wavelength,
\begin{equation}
r-r\cos\theta_c \lesssim \lambda_{\rm FRB} \ \Rightarrow \ \frac{1}{2}\theta_c^2 r \lesssim \lambda_{\rm FRB},
\end{equation}
which gives
\begin{equation}
\begin{aligned}
\theta_c\lesssim \left(\frac{2\lambda_{\rm FRB}}{r}\right)^{1/2}&\simeq
(10^{-5} \ {\rm rad}) \ \nu_{\rm FRB,9}^{-1/2}\Gamma_2^{-1}\Delta t_{\rm FRB,-3}^{-1/2}.
\end{aligned}
\end{equation}
Waves from different regions on the shock front therefore superpose incoherently when the emission originates at $\theta>\theta_c$.
We note that the characteristic radiation cone $\sim 1/\Gamma$ is much wider than $\theta_c$, which implies that waves emitted from different regions of the shock front lose phase coherence and are therefore superposed incoherently.

We integrate the emission over the spherical shock surface as follows.
The radiation originates from a thin, relativistically expanding spherical shell.
Each emitting surface element moves with its own radial velocity direction.
The electromagnetic fields measured in the downstream frame must be transformed individually for each surface element into the observer frame before summing their contributions.
For a surface element at an angular separation $\psi_{\rm obs}$ from the LOS, the local plasma velocity (which is radial) makes an angle $\psi_{\rm obs}$ with the LOS in the observer frame. 
The corresponding viewing angle $\theta_v$ in the downstream frame is obtained through relativistic aberration (see Appendix~\ref{app:EM}).

We perform a Lorentz transformation of the electromagnetic fields with velocity $\beta c$ along the local radial direction, where $\beta=(1-\Gamma^{-2})^{1/2}$, followed by a projection of the transformed electric field onto the plane of the sky (see Appendix \ref{app:EM} for detailed derivations).
After transforming the fields to the observer frame, the radiation from different surface elements superposes incoherently, yielding the surface-integrated Stokes parameters
\begin{equation}
I_{\rm tot}=\sum_i I_i, \ Q_{\rm tot}=\sum_i Q_i, \ U_{\rm tot}=\sum_i U_i, \ V_{\rm tot}=\sum_i V_i,
\end{equation}
where the subscript $i$ labels the individual emitting surface elements on the spherical shock front, and $I_i$, $Q_i$, $U_i$, and $V_i$ are the Stokes parameters of each surface element after transformation to the observer frame.

\begin{figure*}
\begin{center}
\begin{tabular}{ll}
\resizebox{86mm}{!}{\includegraphics[]{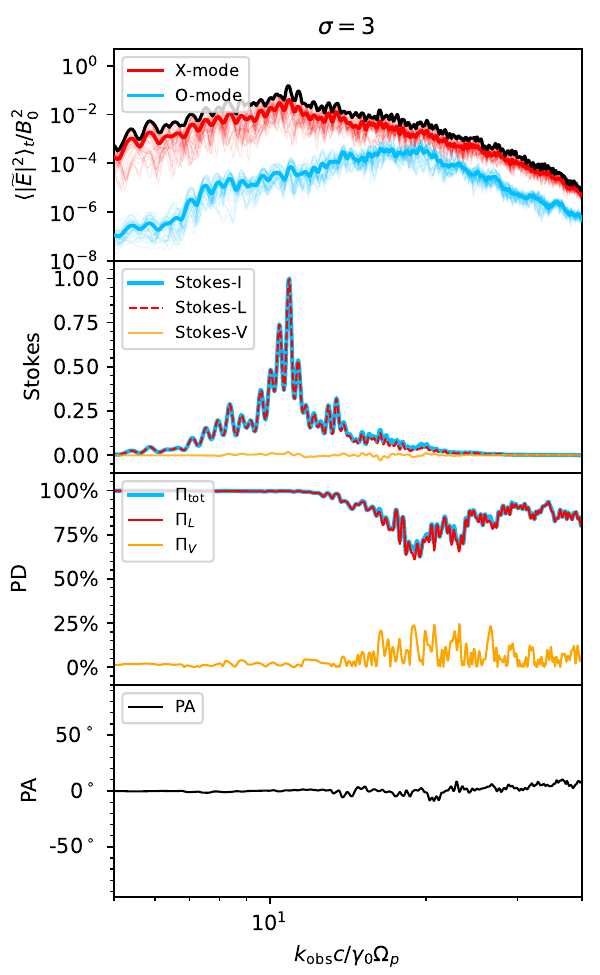}}&
\resizebox{86mm}{!}{\includegraphics[]{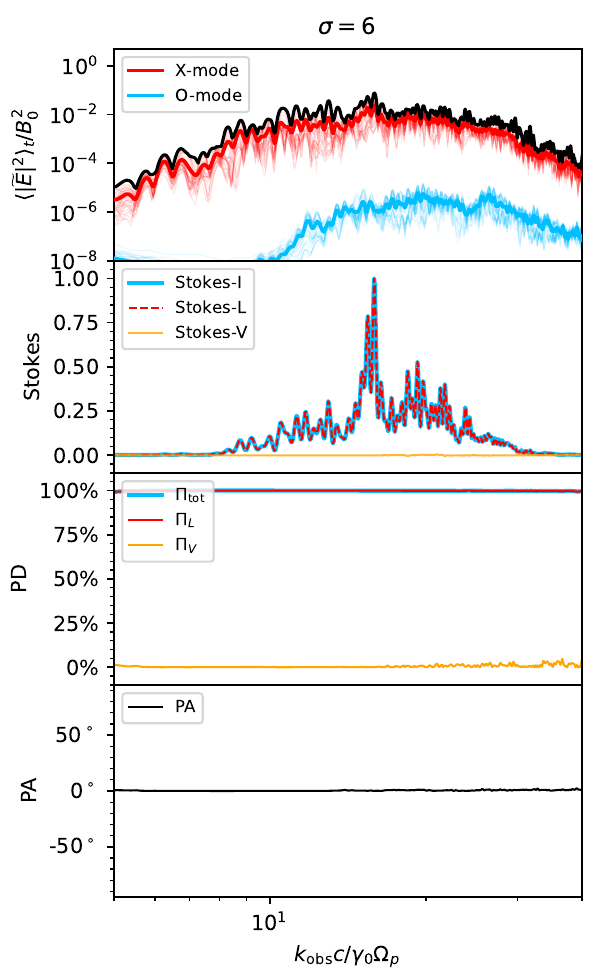}}
\end{tabular}
\caption{Viewing-angle-integrated spectra, Stokes parameters, PD, and PA as a function of $k_{\rm obs}c/\gamma_0\Omega_p$ for $\sigma=3$ and $\sigma=6$ in the observer frame. 
The faint curves in the top panels show the individual snapshots over $1500\leq\Omega_p t\leq2300$, the thick red and blue curves show the time-averaged X- and O-mode spectra over the same interval, respectively.
The black curve shows the maximum combined X- and O-mode power at each wavenumber over the same time interval.
The Stokes parameters are averaged over time before calculating the polarization degrees and PA.
}
\label{fig:integration}
\end{center}
\end{figure*}

\subsection{Radiation spectra}

The first panel of Figure~\ref{fig:integration} shows the shock-integrated spectra of X- and O-modes as a function of $k_{\rm obs}c/\gamma_0\Omega_p$.
To characterize the bandwidth in a manner consistent with observational analyses, we fit the envelope of the integrated combined X- and O-mode spectra with a Gaussian profile and define the fractional bandwidth as $\Delta\nu/\nu_0={\rm FWHM}/\nu_0$, where $\nu_0$ is the central frequency of the fitted Gaussian.
For $\sigma=3$, the fractional bandwidth\footnote{An analytical estimate by \citet{KQZ2024}, assuming monochromatic emission from each local shock patch, gives a fractional bandwidth of $\Delta\nu/\nu_0\approx0.58$ due to the high-latitude effect. In our PIC simulations, each local shock patch has a finite intrinsic spectral width and contains multiple narrow spectral features.} increases from $\Delta\nu/\nu_0\simeq0.41$ for the single-patch spectrum to $\simeq0.57$ after the viewing-angle integration, and for $\sigma=6$ it increases from $\simeq0.74$ to $\simeq0.82$ (see Appendix~\ref{app:bandwidth} and Figure~\ref{fig:bandwidth_methods}).
The bandwidth is broader than the single-patch spectra because emission from higher latitudes experiences different Doppler factors, spreading the observed frequencies.

\subsection{Polarization properties}

The shock-integrated radiation is highly linearly polarized, with the PA remaining nearly constant around the spectral peak due to the dominance of the X-mode. 
Although the incoherent superposition of emission from different latitudes introduces some depolarization at frequencies higher than the peak, the overall PD remains high. 
In contrast, for a single shock patch at a given snapshot, the total polarization degree is unity by construction.
In addition, a non-zero level of circular polarization is present. 
Although the PA is essentially flat near the peak frequency, it shows modest variations of $\sim \pm 10^\circ$ at higher $k_{\rm obs}$.

\section{Spectral bandwidth of a relativistic magnetized spherical shock}\label{sec:analytical}

In this section, we provide an estimate of the spectral broadening caused solely by the high-latitude effect.
We consider a thin spherical shock expanding relativistically with a constant Lorentz factor $\Gamma$.
We assume that the emission is monochromatic in the comoving frame of the shock, with a characteristic frequency $\nu'_0$.
The specific intensity in the comoving frame is
\begin{equation}
I'_{\nu'}=I'_0 \delta(\nu'-\nu'_0),
\end{equation}
where $I'_0$ is assumed to be independent of emission time. 
We further assume that $\Gamma$, $\nu'_0$, and $I'_{\nu'}$ remain constant over time.
Here, $\theta$ is the angle measured relative to the local shock normal.
In the upstream frame, the emission cone is restricted to a maximum angle \citep{Babul&Sironi2020}
\begin{equation}
\theta_{\max}=\frac{A}{\Gamma\sqrt{\sigma}},
\label{eq:theta_max}
\end{equation}
where $A$ is a dimensionless factor of order unity.
The observed frequency can be written as $\nu(\theta)={\cal D}(\theta)\nu'_0$ and the maximum observed frequency is $\nu_0=\nu(0)$.
The Lorentz invariant $I_\nu/\nu^3$ gives $I_\nu={\cal D}^3 I'_{\nu'}$.
For monochromatic emission, the intensity can be written as
\begin{equation}
I_\nu={\cal D}^3 I'_0\delta\left(\frac{\nu}{{\cal D}}-\nu'_0\right).
\label{eq:I_nu}
\end{equation}
The contribution to the observed flux from an annulus between $\theta$ and $\theta+d\theta$ is proportional to $dF_\nu\propto I_\nu\sin\theta d\theta$.
Thus the flux can be calculated as
\begin{equation}
F_\nu\propto\int{\cal D}^3\delta\left(\frac{\nu}{{\cal D}}-\nu'_0\right)\sin\theta d\theta\propto\nu^2.
\label{eq:Fnu_integral}
\end{equation}
The normalized instantaneous spectrum at a fixed observer time is
\begin{equation}
\frac{F_\nu}{F_{\nu,0}}=\left(\frac{\nu}{\nu_0}\right)^2, \ \nu_{\rm min}\leq\nu\leq\nu_0,
\label{eq:Fnu_finl}
\end{equation}
where $F_{\nu,0}$ is the specific flux evaluated at the maximum observed frequency $\nu_0$.
The minimum observed frequency corresponds to the edge of the emitting region, i.e. $\theta=\theta_{\max}=A/(\Gamma\sqrt{\sigma})$. We have
\begin{equation}
\frac{\nu_{\min}}{\nu_0}=
\frac{1-\beta}
{1-\beta\cos\left(\dfrac{A}{\Gamma\sqrt{\sigma}}\right)}\simeq\frac{\sigma}{\sigma+A^2},
\label{eq:nu_min}
\end{equation}
where we have assumed $\Gamma\gg 1$ and $A/(\Gamma\sqrt{\sigma})\ll 1$.
We define $\nu_{1/2}$ as the frequency at which the specific flux drops to half of its peak value $F_{\nu,0}$, i.e.
\begin{equation}
F_{\nu_{1/2}}=\frac{1}{2}F_{\nu,0}, \ \frac{\nu_{1/2}}{\nu_0}=\frac{1}{\sqrt{2}}.
\end{equation}
Thus the FWHM fractional bandwidth is
\begin{equation}
\frac{\Delta\nu}{\nu_0}=2(\nu_0-\nu_{1/2})/\nu_0\simeq 0.58,
\end{equation}
which is consistent with \cite{KQZ2024}.
For the half-maximum point to lie within the emission cone, we require
\begin{equation}
\nu_{\min}\leq\nu_{1/2} \ \Rightarrow \ \frac{\sigma}{\sigma+A^2}\leq\frac{1}{\sqrt{2}}.
\end{equation}
We can define a critical magnetization $\sigma_{\rm crit}$ such that
\begin{equation}
\sigma_{\rm crit}\equiv\frac{A^2}{\sqrt{2}-1}\simeq2.4 A^2.
\label{eq:sigma_crit}
\end{equation}

For $\sigma\leq\sigma_{\rm crit}$, we conclude that the FWHM spectral bandwidth is independent of magnetization.
For $\sigma>\sigma_{\rm crit}$, the angular boundary is reached before
the spectrum decreases to half its peak value. For a relativistic magnetized shock with $\Gamma\gg1$ and $A/(\Gamma\sqrt{\sigma})\ll1$, spectral power sharply drops at
\begin{equation}
\frac{\nu_{\min}}{\nu_0}\simeq\frac{\sigma}{\sigma+A^2},
\end{equation}
where the flux immediately before the cutoff is ${F_{\nu_{\min}}}/{F_{\nu,0}}=[{\sigma}/{(\sigma+A^2)}]^2>1/2$.
The FWHM fractional bandwidth can be calculated as
\begin{equation}
\frac{\Delta\nu}{\nu_0}=\frac{2A^2}{\sigma+A^2}\sim \frac{2A^2}{\sigma} \quad {\rm for}~\sigma\gg 1.
\end{equation}

To summarize the main results of this section, we study the fractional bandwidth of intrinsically monochromatic emission from a relativistic magnetized spherical shock and find that the resulting bandwidth depends on the upstream magnetization.
For $\sigma>\sigma_{\rm crit}$, the fractional bandwidth decreases with increasing magnetization and scales as $\Delta\nu/\nu_0\propto\sigma^{-1}$ in the limit of $\Gamma\gg1$ and $\sigma\gg1$.
Thus, we conclude that higher magnetization leads to a narrower observed spectrum.
This trend applies to the geometric broadening of intrinsically monochromatic emission and does not imply that the full PIC spectrum must become narrower with increasing magnetization, since the local shock emission has some intrinsic finite bandwidth.
The high-latitude effect can still broaden an intrinsically finite-bandwidth spectrum, as demonstrated by the viewing-angle-integrated spectra in Section~\ref{sec:integrated}.
At the same time, higher magnetization shocks are also expected to produce precursor emission increasingly dominated by the X-mode, resulting in high linear polarization degrees and flat PA, while the emission efficiency decreases with increasing magnetization \citep{Plotnikov&Sironi2019,Sironi2021}.

There are two important caveats.
First, the maximum emission angle adopted above is determined by requiring that the precursor wave group velocity along the shock normal exceeds the shock velocity.
Waves that do not satisfy this condition cannot be measured in our simulations, since they cannot reach the upstream region $5c/\Omega_p < x-x_{\rm sh} < 35c/\Omega_p$ ahead of the shock front used in our analysis.
However, whether these waves are ultimately damped or eventually escape ahead of a decelerating shock remains uncertain.
Second, we have assumed that the characteristic frequency and the intrinsic emission intensity remain constant during the shock propagation. 
The intrinsic temporal evolution of the emission frequency and the emitted flux will introduce additional spectral evolution beyond the geometric high-latitude effect that we have considered here.

\section{Conclusions and Discussions}\label{sec:conclusion}

In this work, we have studied the observable spectra and polarization signatures of the synchrotron maser emission from relativistic magnetized pair-plasma shocks. 
Using 3D particle-in-cell simulations, we first calculate the precursor emission from a plane-parallel perpendicular shock, neglecting radiative cooling. 
Then we project the simulated emission onto an arbitrary LOS and apply the planar shock-patch results to a spherical shock front to investigate the effects of different viewing angles and high-latitude emission on the observed radiation.
The main findings of our paper are:
\begin{itemize}
\item {Polarization properties}:
At the peak frequency of the synchrotron maser emission, both the total and linear polarization degrees are $\sim 100\%$. 
We conclude that this emission mechanism can reproduce the extremely high PD of many FRBs.
For a single planar shock patch, when the LOS is aligned with the shock propagation direction, the emission is dominated by the X-mode. 
O-mode emission is required to generate circular polarization, which can arise at off-axis viewing angles. 
However, the globally integrated emission, resulting from incoherent superposition over different regions of a spherical shock, is still expected to be dominated by linear polarization. 
The PA is expected to remain nearly constant around the spectral peak because of the dominance of the X-mode.
At frequencies above the spectral peak, where the emission is fainter, the increasing relative contribution of the O-mode can produce significant circular polarization.
\item {Spectral properties}:
Most repeating FRBs exhibit extremely narrow spectra, with $\Delta\nu/\nu_0 \sim 0.1 - 0.2$ (e.g., FRB 20201124A, FRB 20220912A, FRB 20230607A, and FRB 20240114A). In contrast, non-repeating FRBs show statistically broader spectra \citep{Pleunis2021, CHIMECat2}.
The radiation spectra of a single planar shock patch have finite intrinsic bandwidths and exhibit much narrower spectral spikes.
In Section~\ref{sec:integrated}, we showed that the high-latitude effect further broadens the synchrotron maser spectrum. The Gaussian-envelope fractional bandwidth increases from $\Delta\nu/\nu_0\simeq0.41$ to 0.57 for $\sigma=3$, and from $\simeq0.74$ to 0.82 for $\sigma=6$ after viewing-angle integration (Appendix~\ref{app:bandwidth}).
In Section~\ref{sec:analytical}, our analytical calculation further shows that, in the extreme limit of intrinsically monochromatic emission, the spectral broadening induced by the high-latitude effect decreases as $\Delta\nu/\nu_0\propto\sigma^{-1}$ in the limit of $\Gamma\gg1$ and $\sigma\gg1$. Therefore, at high magnetizations, the observed spectral bandwidth is expected to be determined primarily by the intrinsic bandwidth of the emission of a single shock patch.
For a sensitivity-limited observation in which only the brightest spectral components are detectable, the apparent fractional bandwidths can be much narrower, with $[\Delta\nu/\nu]_{\rm peak}\simeq0.06$ and $0.05$ for $\sigma=3$ and $6$, respectively.
\item  Implications for FRBs:
For the magnetizations considered in this work ($\sigma=3$ and 6), the Gaussian-envelope bandwidths, together with the high linear polarization degrees and nearly flat PAs, are compatible with the properties of some non-repeating FRBs.
If only the brightest spectral components can be detectable due to the limited telescope sensitivity, the apparent fractional bandwidth can be as narrow as $\Delta\nu/\nu\sim0.05-0.06$.
However, in this case, the detectable emission is dominated by the X-mode and is expected to be highly linearly polarized with a nearly flat PA.
This differs from known narrow-band repeating FRBs that can also exhibit strong circular polarization, PA swings, and abrupt PA jumps.
Therefore, the joint observed spectral and polarization properties provide a stringent test of the synchrotron maser shock model.
\end{itemize}

Our simulations do not include the radiation-reaction force, and quantifying its impact on the integrated radiation spectrum remains an open question.
The radiation reaction may become important for GHz emission from extremely relativistic shocks with upstream Lorentz factors $\gamma_0\sim10^4-10^6$ \citep{ZhangYu2026}, well above the Lorentz factor considered in our simulations.
2D PIC simulations of monster shocks within a dipolar magnetosphere show that the shock front can be deformed from a spherical shape \citep{Bernardi2025}. 
For our calculation, spherical symmetry is only required over the angular extent of the emitting region $\sim 1/(\Gamma\sqrt{\sigma})$.
Whether the relativistic magnetized shocks outside the magnetosphere remain approximately spherical on this angular scale requires further investigation.
Another important direction for future investigations is how upstream plasma fluctuations on small scales affect the stability and efficiency of the synchrotron maser mechanism \citep[see e.g.,][]{Demidem2023,Bresci2023}.

\section*{Acknowledgements}
Y.Q. and J.N. are supported by the Research Council of Finland Centre of Excellence in Neutron-Star Physics (project 374063).
J.N. is additionally supported by an ERC grant (ILLUMINATOR, 101114623).
The views and opinions expressed are however those of the authors only and do not necessarily reflect those of the European Union or the European Research Council. 
Neither the European Union nor the granting authority can be held responsible for them. 
L.S. acknowledges support from the DOE Early Career Award DE-SC0023015, NASA ATP 80NSSC24K1826, NSF AST-2307202, NSF PHY-2206609 (the Multimessenger Plasma Physics Center, MPPC), and the Simons Foundation (MP-SCMPS-00001470).
This work is supported by the NASA ATP Grant NNH23ZDA001N.

\appendix
\renewcommand{\theequation}{\Alph{section}\arabic{equation}}
\makeatletter
\@addtoreset{equation}{section}
\makeatother

\section{Polarization measurements for arbitrary viewing angles}\label{app:polarization}

We describe the method for measuring the Stokes parameters and polarization degree from the PIC simulation data.
At a given simulation time, the electric fields are
\begin{equation}
\vec E(x,y,z)=E_x\nvec{x}+E_y\nvec{y}+E_z\nvec{z}.
\end{equation}
We apply a Hamming window along the non-periodic $x$-direction and then apply a discrete Fourier transform to obtain
\begin{equation}
\vec{\widetilde{E}}(k_x,k_y,k_z)=\widetilde{E}_x\nvec{x}+\widetilde{E}_y\nvec{y}+\widetilde{E}_z\nvec{z}.
\end{equation}
Let $\nvec{n}$ denote the unit vector along the LOS. The complex electric field in the plane perpendicular to $\nvec{n}$ is
\begin{equation}
\begin{aligned}
\vec{\widetilde{E}}_\perp&=\vec{\widetilde{E}}-(\vec{\widetilde{E}}\cdot\nvec{n})\nvec{n}=\vec{\widetilde{E}}_1+\vec{\widetilde{E}}_2,
\end{aligned}
\end{equation}
where $\widetilde{E}_1$ and $\widetilde{E}_2$ are the complex amplitudes along two orthogonal polarization basis vectors in the plane perpendicular to the LOS.
Without loss of generality, we set the LOS unit vector to
\begin{equation}
\nvec{n}=\sin\theta_v\cos\phi_v\nvec{x}+\sin\theta_v\sin\phi_v\nvec{y}+\cos\theta_v\nvec{z},
\end{equation}
where $\theta_v$ and $\phi_v$ are the viewing angle and azimuthal angle, respectively.
The unit vector perpendicular to the LOS and to the upstream background magnetic field is
\begin{equation}
\nvec{\zeta}=\frac{\nvec{n}\times\nvec{B}_{\rm bg}}{|\nvec{n}\times\nvec{B}_{\rm bg}|}=\sin\phi_v\nvec{x}-\cos\phi_v\nvec{y}.
\end{equation}
Expanding the perpendicular electric field components explicitly gives
\begin{equation}
\begin{aligned}
\vec{\widetilde{E}}_\perp&=[\widetilde{E}_x-\cos\phi_v\sin\theta_v(\widetilde{E}_z\cos\theta_v+\widetilde{E}_x\cos\phi_v\sin\theta_v\\
&+\widetilde{E}_y\sin\phi_v\sin\theta_v)]\nvec{x}+[\widetilde{E}_y-\sin\phi_v\sin\theta_v(\widetilde{E}_z\cos\theta_v\\
&+\widetilde{E}_x\cos\phi_v\sin\theta_v+\widetilde{E}_y\sin\phi_v\sin\theta_v)]\nvec{y}+[\widetilde{E}_z-\cos\theta_v\\
&\times(\widetilde{E}_z\cos\theta_v+\widetilde{E}_x\cos\phi_v\sin\theta_v+\widetilde{E}_y\sin\phi_v\sin\theta_v)]\nvec{z}.
\end{aligned}
\end{equation}
Projecting $\vec{\widetilde{E}}_\perp$ onto $\nvec{\zeta}$ yields the first eigen-mode,
\begin{equation}
\begin{aligned}
\vec{\widetilde{E}}_1=(\vec{\widetilde{E}}_\perp\cdot\nvec{\zeta})\nvec{\zeta}&=(\widetilde{E}_x\sin\phi_v-\widetilde{E}_y\cos\phi_v)\sin\phi_v\nvec{x}\\
&+(\widetilde{E}_y\cos\phi_v-\widetilde{E}_x\sin\phi_v)\cos\phi_v\nvec{y}.
\end{aligned}
\end{equation}
The corresponding complex amplitude of $\vec{\widetilde{E}}_1$ is
\begin{equation}
{\widetilde{E}}_1=\widetilde{E}_x\sin\phi_v-\widetilde{E}_y\cos\phi_v.
\end{equation}
The complex amplitude of the second mode perpendicular to the first can be calculated as
\begin{equation}
\begin{aligned}
{\widetilde{E}}_2&=\vec{\widetilde{E}}_\perp\cdot(\nvec{\zeta}\times\nvec{n})\\
&=({\widetilde{E}}_x\cos\phi_v+{\widetilde{E}}_y\sin\phi_v)\cos\theta_v-{\widetilde{E}}_z\sin\theta_v.
\end{aligned}
\end{equation}
For a given viewing direction $(\theta_v,\phi_v)$, we extract the radiation spectra along the corresponding ray in wave-vector space,
\begin{equation}
\boldsymbol{k}=k\hat{\boldsymbol n}=k(\sin\theta_v\cos\phi_v,\sin\theta_v\sin\phi_v,\cos\theta_v).
\end{equation}
The Fourier-space electric field components at these wave vectors are obtained by interpolation on the Cartesian $(k_x,k_y,k_z)$ grid.
The resulting complex amplitudes $\widetilde E_1(k,\theta_v,\phi_v)$ and $\widetilde E_2(k,\theta_v,\phi_v)$ are then used to calculate the X- and O-mode spectra and the Stokes parameters at each wavenumber (Section~\ref{sec:individual}).

\section{Lorentz transformations and integration over the spherical shock}
\label{app:EM}

We describe the transformation of the electromagnetic waves measured in the downstream frame to the observer frame and their integration over the surface of an expanding spherical shock.
Each PIC simulation represents a local planar patch of the shock surface. 
Throughout this Appendix, quantities without a frame subscript are measured in the downstream (simulation) frame, consistent with the convention adopted in the main text, and quantities measured in the observer frame are denoted by the subscript "obs".
For a given local patch, the Fourier-space wave fields are
\begin{equation}
\widetilde{\vec E}=(\widetilde E_x,\widetilde E_y,\widetilde E_z), \ \widetilde{\vec B}=(\widetilde B_x,\widetilde B_y,\widetilde B_z).
\end{equation}
For a surface element with local radial velocity $\vec \beta c$, the wave fields are transformed from the downstream frame to the observer frame using the Lorentz transformation \citep{Jackson1998}
\begin{equation}
\begin{aligned}
\widetilde{\vec E}_{\rm obs}=&\Gamma\left(\widetilde{\vec E}-\vec{\beta}\times\widetilde{\vec B}\right)-\frac{\Gamma^2}{\Gamma+1}\vec{\beta}\left(\vec{\beta}\cdot\widetilde{\vec E}\right),\\
\widetilde{\vec B}_{\rm obs}=&\Gamma\left(\widetilde{\vec B}+\vec{\beta}\times\widetilde{\vec E}\right)-\frac{\Gamma^2}{\Gamma+1}\vec{\beta}\left(\vec{\beta}\cdot\widetilde{\vec B}\right),
\end{aligned}
\label{eq:lorentz_fields}
\end{equation}
where $\Gamma=(1-\beta^2)^{-1/2}$.
The wave four-vector is transformed consistently with the electromagnetic fields. 
The frequency and wave vector in the observer frame can be written as
\begin{equation}
\omega_{\rm obs}=\Gamma\left(\omega+ c\vec{\beta}\cdot\vec{k}\right),
\end{equation}
and
\begin{equation}
\vec{k}_{\rm obs}=\vec{k}+\left[\frac{\Gamma-1}{\beta^2}(\vec{\beta}\cdot\vec{k})+\Gamma\frac{\omega}{c}\right]\vec{\beta}.
\label{eq:lorentz_k}
\end{equation}
The precursor waves considered here are approximately vacuum electromagnetic waves (Section~\ref{sec:dispersion}).

For spherical integration, we consider that $\psi_{\rm obs}$ denotes the angle between the local radial velocity and the LOS in the observer frame.
In the local basis of each shock patch, we choose
$\hat{\boldsymbol x}_{\rm loc}$ along the radial velocity and $\hat{\boldsymbol z}_{\rm loc}$ along the background magnetic field, then we have $\hat{\boldsymbol x}_{\rm loc}\perp\hat{\boldsymbol z}_{\rm loc}$ and we define $\hat{\boldsymbol n}_{\rm obs}$ as the LOS in the observer frame, i.e., $\hat{\boldsymbol n}_{\rm obs}=\cos\psi_{\rm obs}\hat{\boldsymbol x}_{\rm loc}+\sin\psi_{\rm obs}\hat{\boldsymbol z}_{\rm loc}$.
The radiation propagation direction associated with the surface element is transformed from the observer frame to the downstream frame using relativistic aberration
\begin{equation}
\hat{\boldsymbol n}_x=\frac{\cos\psi_{\rm obs}-\beta}{1-\beta\cos\psi_{\rm obs}}, \ \hat{\boldsymbol n}_z=\frac{\sin\psi_{\rm obs}}{\Gamma(1-\beta\cos\psi_{\rm obs})}.
\label{eq:aberration}
\end{equation}
The viewing angle is the angle between the aberrated LOS and the local background magnetic field,
\begin{equation}
\cos\theta_v=\hat{\boldsymbol n}\cdot\hat{\boldsymbol z}_{\rm loc}=\frac{\sin\psi_{\rm obs}}{\Gamma(1-\beta\cos\psi_{\rm obs})}.
\label{eq:theta_mapping}
\end{equation}
Thus each latitude $\psi_{\rm obs}$ on the spherical shell maps, through relativistic aberration, onto a corresponding viewing angle $\theta_v$ in the downstream frame.
The integration covers the full range of aberrated viewing angles $0\leq\theta_v\leq90^\circ$ in the downstream frame.
The assumed toroidal background magnetic field is perpendicular to the local radial direction at every point on the spherical shock. Thus, when mapping the planar PIC calculation to different surface elements, both the local velocity direction and the local background field direction rotate together, i.e. $\vec B_{0}\cdot\hat{\vec r}=0$ for every patch.
Since the local background magnetic field is perpendicular to the boost direction, its magnitude transforms as $B_{0,\rm obs}=\Gamma B_0$, where $B_0$ is the background field in the downstream frame.
Consequently, the transformed radiation fields are normalized by $B_{0,\rm obs}$ when calculating the radiation spectra in the observer frame (Figure~\ref{fig:integration}).
The radiation field of each patch is projected onto a common basis in the observer's plane of the sky.
We define $\hat{\vec e}_{\rm 1,obs}=\nvec{\zeta}_{\rm obs}, \ \hat{\vec e}_{\rm 2,obs}=\hat{\vec n}\times\hat{\vec e}_{\rm 1,obs}$ to obtain the two complex electric field amplitudes as
\begin{equation}
\widetilde E_{1,\rm obs}=\widetilde{\vec E}_{\rm obs}\cdot
\hat{\vec e}_{\rm 1,obs}, \ \widetilde E_{2,\rm obs}=\widetilde{\vec E}_{\rm obs}\cdot\hat{\vec e}_{\rm 2,obs}.
\end{equation}
The Stokes parameters of each surface element are then calculated
from $\widetilde E_{1,\rm obs}$ and $\widetilde E_{2,\rm obs}$ using Equation~(\ref{eq:Stokes}).
Before summing the contributions from different surface elements,
the Doppler-shifted fields are interpolated onto a common wavenumber grid in the observer frame.
Radiation from spatially separated surface elements is summed incoherently, as discussed in Section~\ref{sec:integrated}. 
The polarization degrees and polarization angle of the integrated radiation are finally evaluated from the summed Stokes parameters.

\section{Characterization of the spectral bandwidth}
\label{app:bandwidth}

We characterize the spectral width of the shock-integrated total emission using a Gaussian fit in Section~\ref{sec:integrated}. 
Here we describe the procedure in more detail and also consider an alternative measure that is more sensitive to the brightest spectral components.

\begin{figure*}
\begin{center}
\setlength{\tabcolsep}{-6pt}
\begin{tabular}{ll}
\resizebox{95mm}{!}{\includegraphics[]{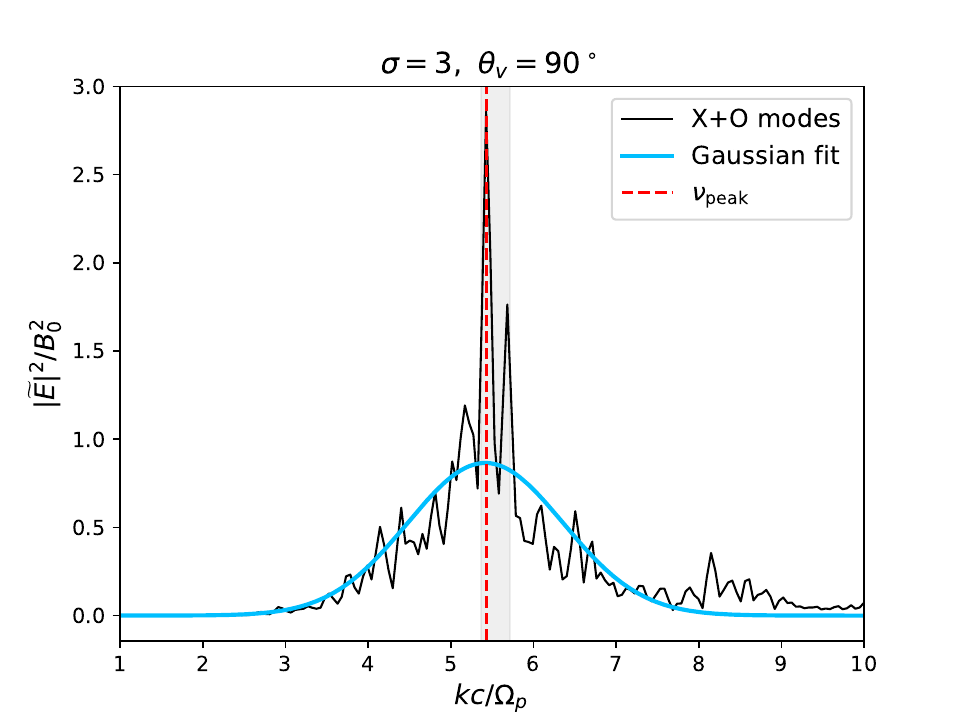}}&
\resizebox{95mm}{!}{\includegraphics[]{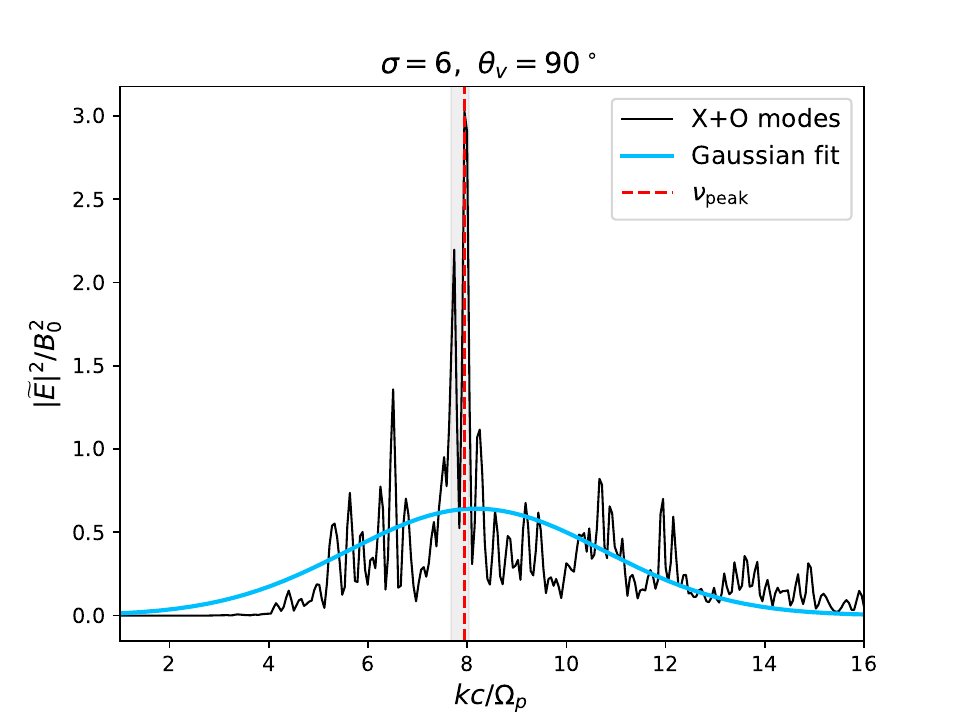}}\\
\resizebox{95mm}{!}{\includegraphics[]{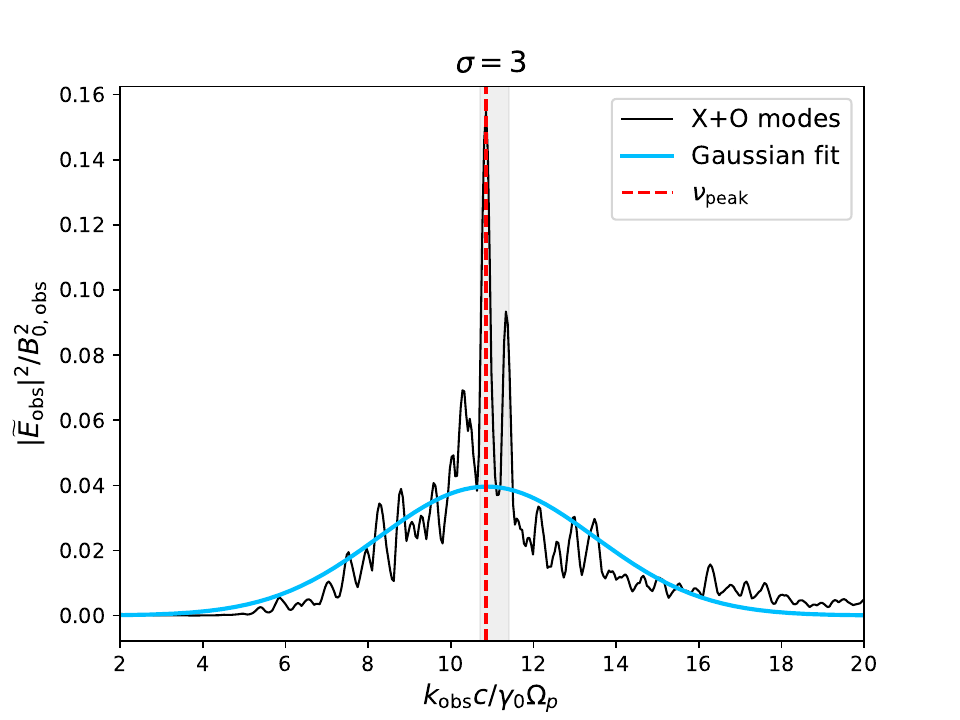}}&
\resizebox{95mm}{!}{\includegraphics[]{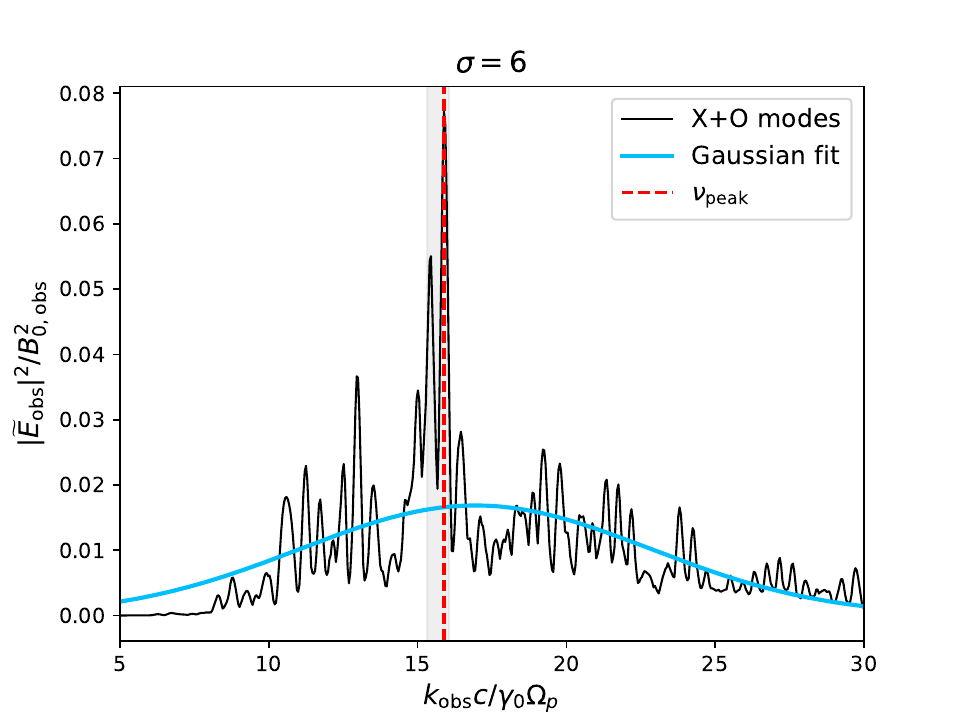}}
\end{tabular}
\caption{Comparison of the spectral bandwidths of the single-patch and viewing-angle-integrated spectra.
The upper panels show the single-patch spectra at $\theta_v=90^\circ$, the lower panels show the corresponding viewing-angle-integrated spectra, for $\sigma=3$ (left) and $\sigma=6$ (right).
The black curves show the combined spectral power of the X- and O-modes, and the blue curves show Gaussian fits to the broad spectral envelope.
The red dashed lines mark the peak frequencies.
The grey regions indicate the high-flux spectral interval adopted for the alternative sensitivity-limited estimate.
}
\label{fig:bandwidth_methods}
\end{center}
\end{figure*}

For a direct comparison with the fractional bandwidths inferred from FRB observations, we characterize the overall spectral envelope using a Gaussian function. 
This method follows the commonly adopted observational characterization of narrow band FRB spectra, in which the central frequency and bandwidth are obtained from a Gaussian fit to the burst spectra \citep[e.g.,][]{Law2017,Aggarwal2021,Nimmo2023,ZhangYK2023,Gopinath2024,ZhouDJ2025,ZhangLX2026}.
Specifically, we fit both the single-patch and angle-integrated combined X- and O-mode spectra with
\begin{equation}
y(\nu)=A_0\exp\left[-\frac{(\nu-\nu_0)^2}{2\sigma_\nu^2}\right],
\label{eq:gaussian_bandwidth}
\end{equation}
where $A_0$ is the normalization, $\nu_0$ is the fitted central frequency, and $\sigma_\nu$ is the standard deviation of the Gaussian.
The spectral bandwidth is defined as the FWHM $\Delta \nu=2\sqrt{2\ln2}\sigma_\nu$ and the fractional bandwidth can be written as
\begin{equation}
\frac{\Delta\nu}{\nu_0}=\frac{2\sqrt{2\ln2}\sigma_\nu}{\nu_0}.
\label{eq:relative_bandwidth}
\end{equation}
Figure~\ref{fig:bandwidth_methods} illustrates this procedure for the single-patch and angle-integrated combined X- and O-mode spectra for $\sigma=3$ and $\sigma=6$.
Although the spectrum contains narrow spectral substructures, the Gaussian fit is intended to characterize the broad envelope rather than individual spectral spikes. We obtain
\begin{equation}
\begin{aligned}
&{\text{single-patch:}} \ &&\left[\frac{\Delta\nu}{\nu_0}\right]_{\sigma=3}\approx0.41, \ \left[\frac{\Delta\nu}{\nu_0}\right]_{\sigma=6}\approx0.74.\\
&{\text{angle-integrated:}} \ &&\left[\frac{\Delta\nu}{\nu_0}\right]_{\sigma=3}\approx0.57, \ \left[\frac{\Delta\nu}{\nu_0}\right]_{\sigma=6}\approx0.82.
\end{aligned}
\end{equation}
This definition is useful for comparison with observational analyses, where the fitted Gaussian center and FWHM are commonly used to characterize the central frequency and spectral bandwidth of FRB bursts.

The Gaussian fit above assumes that the broad spectral envelope is detectable. However, a telescope with limited sensitivity may preferentially detect only the brightest portions of the spectra, while the fainter underlying spectral components fall below the detection threshold. 
The simulated spectra contain a number of narrow and bright spectral spikes, thus the apparent observed bandwidth can be smaller than the width inferred from the full spectral envelope.
We introduce an alternative estimate based on the high-flux spectral components. 
We identify the peak spectral intensity at $\nu_{\rm peak}$ and measure the bandwidth $\Delta\nu=\nu_+-\nu_-$, where $\nu_-$ and $\nu_+$ are the lowest and highest frequencies, respectively, at which the spectral intensity reaches half of its peak value.
The corresponding fractional bandwidth is defined as
\begin{equation}
\left[\frac{\Delta\nu}{\nu}\right]_{\rm peak}=\frac{\nu_+-\nu_-}{\nu_{\rm peak}}.
\end{equation}
We obtain
\begin{equation}\label{eq:bandwidth_peak}
\begin{aligned}
&{\text{single-patch:}} \ &&\left[\frac{\Delta\nu}{\nu}\right]_{\rm peak}^{\sigma=3}\approx0.06, \ \left[\frac{\Delta\nu}{\nu}\right]_{\rm peak}^{\sigma=6}\approx0.05.\\
&{\text{angle-integrated:}} \ &&\left[\frac{\Delta\nu}{\nu}\right]_{\rm peak}^{\sigma=3}\approx0.06, \ \left[\frac{\Delta\nu}{\nu}\right]_{\rm peak}^{\sigma=6}\approx0.05.
\end{aligned}
\end{equation}

An even narrower observed bandwidth could arise if the telescope sensitivity is insufficient to detect the full spectral envelope.
In this case, only the brightest part of the spectrum might be detectable above the sensitivity threshold, resulting in an apparent bandwidth substantially narrower than that inferred from the Gaussian envelope. 
Moreover, an individual narrow spectral spike can be even narrower than the estimate in Equation~(\ref{eq:bandwidth_peak}).
The inferred bandwidth would then be substantially narrower than that obtained from the Gaussian fit to the full spectral envelope.
Since the brightest spectral features are strongly dominated by the X-mode, such a detection would also be expected to exhibit a very high degree of linear polarization with $\Pi_L\sim100\%$.
However, only a fraction of the total emitted flux would be detected, leading to a lower observed flux.

\bibliography{example}{}
\bibliographystyle{aasjournal}

\end{document}